%% file: fractional-cosmology.tex
\documentclass[aps,citeautoscript,twocolumn,groupedaddress,amsfonts,amssymb,amsmath,showpacs,showkeys,nofootinbib]{revtex4-2}

\usepackage{eucal}

\usepackage{graphicx}

\usepackage{color}

\usepackage{fancybox}

\usepackage{psfrag}
\usepackage{natbib}
\usepackage{setspace}

\usepackage[dvipsnames]{xcolor}

\input{aas-journals}

\begin{document}

% THIS IS THE TITLE OF THE ARTICLE:

\title{Extending Friedmann equations using fractional derivatives using a Last
Step Modification technique: the case
of a matter dominated accelerated expanding Universe. }

% \author{E. Barrientos$^{1,2}$, S. Mendoza$^1$ and P. Padilla$^2$}
% \email[Emails: ]{ernestobar14@ciencias.unam.mx\\
%                  sergio@astro.unam.mx\\
% 		 pablo@mym.iimas.unam.mx}
% \affiliation{$^1$Instituto de Astronom\'{\i}a, Universidad Nacional
%                  Aut\'onoma de M\'exico, AP 70-264, Ciudad de M\'exico 04510,
% 	         M\'exico.\\
% 	     $^2$Instituto de Investigaciones en Matem\'aticas Aplicadas 
% 	         y en Sistemas, Departamento de Matem\'aticas y
% 		 Mec\'anica, Universidad Nacional Aut\'onoma de M\'exico,
% 		 Ciudad de M\'exico 04510, M\'exico.
%             }

\author{E. Barrientos}
  \email[Email: ]{ernestobar14@ciencias.unam.mx}
\affiliation{
  Instituto de Astronom\'{\i}a, Universidad Nacional
   Aut\'onoma de M\'exico, AP 70-264, Ciudad de M\'exico 04510, M\'exico.
}
\affiliation{
  Instituto de Investigaciones en Matem\'aticas Aplicadas 
    y en Sistemas, Departamento de Matem\'aticas y
    Mec\'anica, Universidad Nacional Aut\'onoma de M\'exico,
    Ciudad de M\'exico 04510, M\'exico.
}

\author{S. Mendoza}
  \email[Email: ]{sergio@astro.unam.mx}
\affiliation{
  Instituto de Astronom\'{\i}a, Universidad Nacional
   Aut\'onoma de M\'exico, AP 70-264, Ciudad de M\'exico 04510, M\'exico.
}

\author{P. Padilla}
  \email[Email: ]{pablo@mym.iimas.unam.mx}
\affiliation{
  Instituto de Investigaciones en Matem\'aticas Aplicadas 
    y en Sistemas, Departamento de Matem\'aticas y
    Mec\'anica, Universidad Nacional Aut\'onoma de M\'exico,
    Ciudad de M\'exico 04510, M\'exico.
}

\date{\today}

\begin{abstract}
  We present a toy model for extending the Friedmann equations of relativistic 
cosmology using fractional derivatives.  
We do this by replacing the integer derivatives, in a few well-known cosmological
results with fractional derivatives leaving their order as a free parameter.  All 
this with the intention  to explain the current 
observed acceleration of the Universe.
We apply the Last Step
Modification technique of fractional calculus to construct some useful 
fractional equations of cosmology.  
The fits of the unknown fractional derivative order and 
the fractional cosmographic parameters to SN~Ia data shows that this simple 
construction can 
explain the current accelerated expansion of the Universe without the
use of a dark energy component with a MOND-like behaviour using 
Milgrom's acceleration constant which sheds light into to the non-necessity of a dark matter 
component as well.
\end{abstract}

% PACS numbers:
% Symmetry does not require PACS.  Just keywords.
% \pacs{3.1416-q}
\keywords{Cosmology; SN~Ia; fractional calculus.}

\maketitle

\section{Introduction}
\label{Introduccion}

  The Hilbert--Einstein field equations, a success of general relativity,
work tremendously well at mass-to-length scales similar to those of the
solar system~\citep[see e.g.][]{Will}, where the gravitational field is
moderately weak \citep{Eddington, Pound, Reasenberg, Anderson, Chandler,Ciufolini,Sitter, Eubanks, Nordtvedt, Nordtvedt2, Nordtvedt3, Schiff, Shapiro}.  
At very large mass-to-length ratios,
where the gravitational field is very strong, the detection of gravitational waves
produced by the interactions of compact objects such as black 
holes and neutron stars has 
shown a remarkable good agreement with the predictions of general relativity.

  When the Newtonian acceleration of a test particle reaches values \(
a_0 \lesssim 10^{-10} \textrm{m}/\textrm{s}^2 \), or equivalently when the mass-to-length
ratios are much smaller than those of the solar system~\citep{mendoza15}, 
the Hilbert--Einstein field
equations cannot fit an enormous amount of astrophysical and cosmological data 
unless: (a) extra dark
matter and/or dark energy components are added or (b) the curvature caused 
by the matter and energy requires extensions or
modifications. In other words, the Hilbert--Einstein field equations do not
correctly predict the observed results at those acceleration scales unless (a) or (b) are adopted.
In what follows, we will only deal with case (b) for the accelerated
expansion of the Universe at the present epoch.

  If the gravitational field equations are to be extended, the first intuitive attempt
consists on assuming a general \( f(R) \) function of the Ricci curvature scalar \( R \) as the 
Lagrangian in the gravitational action.
Despite some $f(R)$ theories have interesting results \citep{staro,Nojiri11,Nojiri08,shamir,Odin20,Nor-Odin20,Nor-Odin-04,Nor-Odin-03}, 
there is currently not a full  $f(R)$ Lagrangian which solves all the shortcomings between observations and the gravitational
theory. Through the years, more general actions have been proposed, using for example 
functions of several scalars built with the Riemann tensor and even ones in which 
couplings between \( R \) and the matter Lagrangian or the 
trace of the energy-momentum tensor~\citep{frlm,Harko1,Harko3,Harko4,harko-lobo-book, Bertolami,barrientos18} are adopted.
These proposals are still being developed and investigated in full, and although interesting
in principle, they have a small general inconvenient: the motion of free particles is not 
necessarily geodesic and as such, a fifth force appears naturally.

  In \citet{bernal11,mendoza13,barrientos16,barrientos18,Barrientos20}, the authors proposed
general gravitational actions with curvature-matter couplings in order to obtain a feasible 
explanation for the MOdified Newtonian Dynamics (MONDian) behavior of gravitational phenomena.  The general conclusions
reached by the works of~\citet{bernal11,mendoza13,Barrientos20} is that it is possible to 
recover the MONDian expression for the acceleration in the regime $a<a_0$ for a pure metric 
$f(R)$  theory provided a non-local action construction. Non-locality in these
attempts is introduced  as an extra scalar field with dimensions of mass that can be 
conveniently thought of as the causally-connected mass to each point in space-time.

The idea of a non-local gravity and its implications at solar system and cosmological scales have 
been recently revisited by the works of~\citet{Mash1,Mash2,Chicone1,Chicone2,Chicone3, Blome}. These theories 
claim that locality, which allows to treat an accelerated observer as an inertial one at each
instant on his proper world-line in special relativity, has its limitations. Since 
the Einstein Equivalence Principle
relates an observer in a local gravitational field with another accelerated one with
no gravitation, those limitations 
are thus extended to gravitational theory. A general characteristic of non-locality is that fields are no longer 
given by their instant (or local) value but 
have a contribution attached to the history of the observer (in general terms the Lagrangian density is not localised and as such is not defined 
as a simple function of the space-time coordinates).  Recent investigations 
have shown that these kind of proposals can simulate dark  matter behaviour~\citep{Maggiore1,HehlM, Foffa}. 
For Tully--Fisher scalings, MOND introduces a gravitational 
potential for a point mass source  
proportional to $\ln(r)$, which flattens rotation
curves.  This is also included in these non-local proposals, but they cannot 
fully explain the proportionality of the MOND force to \( \sqrt{G} \), where $G$ is Newton's
gravitational constant.   For the non-local constructions, the relation 
between the acceleration and the gravitational constant turns out to be linear.

  In the present article, we discuss another possible non-local approach in which fractional calculus is used.
Fractional calculus, although still a curiosity for many, has proved to describe appropriately 
non-local space-time effects in a wide range of applications 
\citep{Podlubny, Handbook1,Handbook2,Handbook3,Handbook4,Handbook6,Handbook7}.  In general terms, 
fractional calculus extends the order of differentiation and integration from the natural to the real numbers (in fact to the complex numbers, but we are only going to work with real values in this work).   In order to simplify
the understanding of fractional calculus to the non-expert, we 
have included in Appendix~\ref{apen1} a simple introduction to fractional calculus.

 The introduction of fractional derivatives in gravity is not a trivial task since there are several 
proposals in order to perform the required generalisations. From a pure 
mathematical point of view, fractional derivatives must induce a somehow fractional geometry 
in such a way that all the geometric entities involved in general relativity e.g.~the connection, 
the covariant derivative, the Riemann tensor and the metric should 
be defined in terms of a fractional derivative order. Such an 
approach is known as the First Step Modification (FSM). A more 
practical and simple way consist in modifying the Hilbert--Einstein field 
equations for a given geometry, 
replacing the covariant derivative order by its analogous fractional derivative without going 
deeper in how such equation can be obtained. This last approach is usually called a  Last Step Modification (LSM). 
In recent years the applicability of both approaches has been studied in gravity at a classical and 
and cosmological level~\citep{shch1,shch2,shch3,shch4,Roberts,Vacaru,Rami}. An intermediate approach is to formulate 
a variational principle for a fractional order action. This latter approach is of particular interest 
for the scientific community, not exclusively of the gravitational area, since a fractional variational 
theory is general enough for applications in different scientific
fields~\citep{Frederico,Balenu07,Nabulsi,Herzallah,Baleanu05,Nabusi13}. 

The introduction of fractional derivatives (with unknown derivative orders, which are to be 
fitted to observations) in the
Friedmann equations  means that either the gravitational Lagrangian 
contains fractional derivatives, or the
order of variation has a non-integer value (or both).  In general terms,
if fractional derivatives are to be allowed in physical equations, then
the order of differentiation should also be a parameter in the action.  In
other words, the principle of least action for field equations should somehow
contain fractional derivatives.  If this is not the case, another possibility is that the
variation of the action includes a fractional
operator.  In a more complicated scenario, both the variation
and the Lagrangian could contain fractional derivatives.  By itself, this 
constitutes a very profound subject outside the scope of this work. In this
article we force the introduction of fractional derivatives into the
standard Friedmann equation of cosmology and see whether we can 
fit the order of derivatives using SN~Ia observations.  
The goal of this
article is to find out whether it would be possible to explain the 
accelerated expansion of the universe without the introduction of  a dark energy
component into the cosmological density budget in the Friedmann equation for a dust flat universe as observed today.  We deal with the dark matter
component by introducing MOND's acceleration constant \( a_0 \) into the dynamics of the 
universe and in principle one would expect that this will force the end result to mean that
a non-baryonic dark matter component is not required.  The end result is that the fractional 
derivative order used to generalise the Friedmann equations turns out to be coherent with the reported value of a full MONDian construction of gravity using fractional derivatives.  
Nevertheless, as it will be discussed in the article,
in general terms with the studies performed we can not completely claim the non-necessity of a
non-baryonic dark matter, but since MONDian effects are introduced into the problem we expect
that component to have a null value.

 The article is organised as follows.  In section~\ref{general-relativity}
we describe a few key results of standard cosmology and cosmographic
parameters.  In Section~\ref{fractional-friedmann} we extend the Friedmann
equations and other relations --including cosmographic parameters-- to their
fractional derivative counterparts.  Since the derivative order is a free 
parameter of the proposal, using these fractional extensions we calibrate 
all free parameters using SN~Ia observations for the accelerated expansion
of the Universe in Section~\ref{cosmography}.  In Section~\ref{sub_resul} we
present our statistical results and finally in Section~\ref{discussion} we
state final remarks of the toy model developed in this article.

\section{Standard cosmology}
\label{general-relativity}

  In this section we briefly summarise a few standard concepts of 
relativistic cosmology that will
be extended to their fractional derivatives counterparts later on.
Many of the results presented in this section
can be found elsewhere~\citep[see e.g.][and references therein]{Peacock,Misner,Longair,Peebles}.  

The Friedmann--Lema$\hat{\textrm{\i}}$tre--Robertson--Walker (FRLW) metric describes an isotropic and homogeneous Universe,  and is given by :

\begin{equation}
    \mathrm{d}s^2=c^2
     \mathrm{d}t^2-a^2(t)\left(\frac{\mathrm{d}r^2}{1-k
     r^2}+r^2 \mathrm{d} \theta^2+r^2 \sin^2\theta \mathrm{d}\varphi^2
     \right),
    \label{FRLW}
\end{equation}

\noindent where \( c \) is the speed of light, \( a(t) \) is is the cosmological scale factor
as a function of the cosmic time \( t \), \( r \) is the radial distance, \( k \) is the curvature
and \( \theta \) and \( \varphi \) are the polar and azimuthal angles respectively.
Substitution of the FLRW metric into the field equations of general relativity with a cosmological constant  \( \Lambda \) 
yields two independent expressions for the time \( 00 \) and radial 
\( 11 \) components respectively:

\begin{eqnarray}
\frac{\dot{a}^2+kc^2}{a^2}=\frac{8\pi G\rho +\Lambda c^2}{3}, 
\label{F1}\\
\frac{\ddot{a}}{a}=-\frac{4\pi G \rho-\Lambda c^2}{3},
\label{F2}
\end{eqnarray}

\noindent for a dust, i.e.~pressure-less Universe. In the previous 
equations $\rho$ represents the matter density.
Using the  standard definition for the Hubble parameter:

\begin{equation}
H(t) := \frac{\dot{a}}{a},
\label{Hubble}
\end{equation}

\noindent equation~\eqref{F1} turns into:

\begin{equation}
    H^2=\frac{8\pi G \rho+\Lambda c^2}{3}-\frac{kc^2}{a^2}.
    \label{FriedH}
\end{equation}

\noindent The right hand side of this equation contains all the information for a late-time Universe's constituents: the curvature \( k \), the matter density \( \rho \) and the 
cosmological constant \( \Lambda \).
The previous equation means that:

\begin{equation}
    1=\Omega_\text{M}+\Omega_\Lambda+\Omega_k,
\label{suma1}
\end{equation}

\noindent where:

\begin{equation}
    \Omega_\text{M} :=\frac{8\pi G \rho}{3H^2}, \quad \Omega_\Lambda :=\frac{c^2\Lambda}{3H^2},
    \quad \Omega_k := -\frac{kc^2}{H^2a^2},
    \label{densities}
\end{equation}

\noindent represent the matter, dark energy and curvature density parameters
respectively. Equation~\eqref{suma1} is a convenient normalisation
for all the energy constituents of the universe, so that their sum equals
one.

We now define a few cosmographic parameters.  To begin with, the 
deceleration parameter:

\begin{equation}
    q := -\frac{1}{H^2}\frac{\ddot{a}}{a},
\label{def_q}
\end{equation}

\noindent can be rewritten as: 

\begin{equation}
    q=\frac{\Omega_\text{M}}{2}-\Omega_\Lambda,
    \label{deceleration}
\end{equation}

\noindent by means of equation~\eqref{F2}.

  Derivating  with respect to cosmic time \( t \) the second Friedmann
equation~\eqref{F2} yields:

\begin{equation}
    \frac{\dddot{a}}{a}-\frac{\ddot{a}\dot{a}}{a^2}=-\frac{4\pi G \dot{\rho}}{3}.
    \label{derivative1}
\end{equation}

\noindent We now introduce another cosmographic parameter, namely the jerk: 

\begin{equation}
  j := \frac{1}{H^3}\frac{\dddot{a}}{a}.   
\label{def_j}
\end{equation}

\noindent In order to express the jerk in terms of the density parameters, we 
use the fact that the covariant divergence of the energy-momentum tensor vanishes,  i.e.~$\nabla_\mu T^{\mu\nu}=0$, 
For a matter-dominated dust Universe, it follows that this relation yields:

\begin{equation}
    \dot{\rho}=-3H\rho.
    \label{timerho}
    \end{equation}
    
\noindent and so $\rho\propto a^{-3}$. Substitution of this last relation into
equation~\eqref{derivative1} yields:

\begin{equation}
    j=\frac{3}{2}\Omega_\text{M}-q,
    \label{j_q}
\end{equation}

\noindent which can be expressed in terms of energy densities only using relation~\eqref{deceleration}:

\begin{equation}
    j=\Omega_\text{M}+\Omega_\Lambda
    \label{jerk}
\end{equation}

  An additional derivative with respect to cosmic time in
equation~\eqref{derivative1} gives the following expression:

\begin{equation}
    \frac{\ddddot{a}}{a}-2\frac{\dddot{a}\dot{a}}{a^2}+2\frac{\ddot{a}\dot{a}^2}{a^3}-\left(\frac{\ddot{a}}{a}\right)^2
    =-\frac{4\pi G \ddot{\rho}}{3}.
    \label{derivative2}
\end{equation}

  The snap cosmographic parameter $s$ is defined as:

\begin{equation}
    s := \frac{1}{H^4}\frac{\ddddot{a}}{a}.
    \label{def_s}
\end{equation}

\noindent  The time derivative of equation~\eqref{timerho} yields 
$\ddot{\rho}=-3(\dot{\rho}H+\rho\dot{H})=-3(-3\rho H^2+\rho\dot{H})$. 
So, using equation~\eqref{derivative2}, together with the definitions of $H$, $q$, $j$
and $\dot{H}=-H^2(1+q)$, the snap parameter can be written as: 

\begin{equation}
    s=-\frac{3}{2}\Omega_\text{M} (4+q)+2j+2q+q^2,
    \label{s_q_j}
\end{equation}

\noindent which in terms of the density parameters is:

\begin{equation}
s=-3\Omega_\text{M}-\frac{1}{2}\Omega_\text{M}^2+\frac{1}{2}\Omega_\text{M}\Omega_\Lambda+\Omega_\Lambda^2.
\label{snap}
\end{equation}

Equations \eqref{deceleration}, \eqref{jerk} and \eqref{snap} can be expressed in terms of the curvature
density parameter $\Omega_k$ \citep{Kun} instead of the dark energy density parameter $\Omega_\Lambda$ 
using equation~\eqref{suma1}. Since the current observations from Planck \citep{planck} strongly suggest that 
we are living in a flat universe $k=0$, we will work with this value in what follows.  Therefore, equation~\eqref{suma1} simplifies to:
$1=\Omega_\text{M}+\Omega_\Lambda$ and so, the value for the jerk parameter 
in general  relativity has a constant unitary value, i.e.~$j=1$.

\section{Fractional Friedmann equation}
\label{fractional-friedmann}

  In what follows, we explore the cosmological consequences of replacing 
the integer time derivatives in the Friedmann equations~\eqref{F1} and~\eqref{F2} 
with fractional derivatives.  The idea is to fit the unknown derivative 
order to SN~Ia observations.  Following this path, we write down
the fractional Friedmann equations as:

\begin{align}
\left(\frac{D^\gamma a}{Dt^\gamma}\right)^2 &=\kappa a^2 \left(\frac{8\pi G\rho +\Lambda c^2}{3}\right),
\label{fraction1}
\\
\frac{D^\gamma}{Dt^\gamma}\left(\frac{D^\gamma a}{Dt^\gamma}\right) &=\kappa a \left(-\frac{4\pi G \rho-\Lambda c^2}{3}\right),
\label{fraction2}
\end{align}

\noindent for a flat Universe.  The constant $\kappa$, with dimensions of 
$\textrm{time}^{2(1-\gamma)}$ has been introduced 
into the fractional Friedmann equations in order to have dimensional coherence.
The left-hand side of equation~\eqref{fraction2} is written as such since in
general $D^\gamma D^\gamma \neq D^{2\gamma}$. 

  Since our target is to work with a Friedmann fractional model with no 
dark matter, we introduce Milgrom's acceleration constant \( a_0 \) as a fundamental
physical quantity for the description of gravitational phenomena at
cosmological scales.  With this and since the velocity of light \( c \) and 
Newton's gravitational constant \( G \) are also fundamental, using 
Buckingham-$\Pi$ theorem for the dimensional  analysis \citep{sedov}, it follows 
that: 

\begin{equation}
  \kappa= A \left(\frac{a_0}{c}\right)^{2(\gamma-1)}.    
\label{kappa}
\end{equation}

 \noindent where $A$ is a dimensionless constant.
 
  Under the idea of fractional orders on the derivative, we can adapt the
cosmographic parameters in a natural way as follows.  
To begin with, we define the fractional Hubble parameter as: 

\begin{equation}
   H^\star := \frac{1}{a}\frac{D^\gamma a}{Dt^\gamma}.
    \label{fracH}
\end{equation}

Since our intention is to express the Friedmann equation in terms of the density parameter, 
we follow the procedures of Section~\ref{general-relativity} and so, dividing 
equation~\eqref{fraction1}  by \( H^2 \) and using the previous definition
it follows that\footnote{\label{foot01} It is important to mention that the use
of the standard definitions for the density parameters in equation~\eqref{densities} implies
that in this extended fractional Friedmann cosmological model, the sum~\eqref{suma1} is no longer
valid.  This essentially occurs because the critical density to ``close'' a pure matter dominated
universe is no longer given by \( 3 H_0^2 / 8 \pi G \).  One can of course redefine the density
parameters in such a way that the sum~\eqref{suma1} holds.  Indeed, if:

\begin{align*}
    \Omega_\text{M}^\star &=   A \left(\frac{a_0}{c}\right)^{2(\gamma-1)}\Omega_\text{M}\left(\frac{H}{H^\star}\right)^2 \\
\intertext{and} 
    \Omega_\Lambda^\star &=  A \left(\frac{a_0}{c}\right)^{2(\gamma-1)}\Omega_\Lambda\left(\frac{H}{H^\star}\right)^2.
\end{align*}

\noindent then:

\begin{displaymath}
  \Omega_\text{M}^\star + \Omega_\Lambda^\star = 1. 
\end{displaymath}

\noindent However, in order to avoid more 
confusion with new extended definitions we decided to keep the standard cosmological definitions 
of equation~\eqref{densities} at the cost of breaking up the validity of relation~\eqref{suma1}.
}: 

\begin{equation}
    \left(\frac{H^\star}{H}\right)^2 = A \left(\frac{a_0}{c}\right)^{2(\gamma-1)}\left(\Omega_\text{M}+\Omega_\Lambda\right).
    \label{intermedio1}
\end{equation}

\subsection{Matter dominated Universe}
In order to compute the term \( H^\star / H \) in the previous expressions, a further assumption about 
the scale factor $a$ must be made. In standard cosmology, to get how the scale factor evolves as function of $t$,
the different constituents of the Universe are treated separately. In this work, we are going to proceed that way.
Thus, from now on we restrict our study to a matter dominated Universe ($\Omega_\Lambda=0$). For this kind of 
Universe, the following ansatz is proposed~\citep[see e.g.][]{Longair,liddle,dodelson}: 

\begin{equation}
  a=a_1 t^n, 
  \label{ansatz}
\end{equation}

\noindent where
\( a_1 \) is a constant, and
using the rules of fractional derivative for a power law given in Appendix~\ref{apen1}, the fractional Hubble 
parameter \(  H^\star \) 
is given by:

\begin{equation}
    H^\star=\frac{\Gamma(n+1)}{\Gamma(n+1-\gamma)}t^{-\gamma},
    \label{Hstar}
\end{equation}

\noindent where the exponent $\gamma$ that appears in
$t^{-\gamma}$ in the previous equation follows standard algebraic rules. 
Also, the standard Hubble parameter for this scale factor 
is: $H=nt^{-1}$. Thus, $H^\star/H$ can be written as:

\begin{equation}
    \frac{H^\star}{H}=\frac{\Gamma(n)}{\Gamma(n+1-\gamma)}t^{1-\gamma}
    =\frac{\Gamma(n)}{\Gamma(n+1-\gamma)}\left(\frac{H}{n}\right)^{\gamma-1}.
    \label{intermedio2}
\end{equation}

\noindent Substitution of this last result into 
equation~\eqref{intermedio1}  yields:

\begin{equation}
    H=B\frac{a_0}{c}\left(\Omega_\text{M}\right)^{1/2(\gamma-1)},
    \label{intermedio3}
\end{equation}

\noindent where the constant $B$ is defined as:

\begin{equation}
    B:=\left[\frac{\Gamma(n+1-\gamma)}{\Gamma(n)}\right]^{1/(\gamma-1)} n
    A^{1/2(\gamma-1)}.
    \label{Bconstant}
\end{equation}

 At this point, we have complete freedom over $A$ and so,
in order to simplify the Hubble parameter
equation~\eqref{intermedio3}, the following choice is made:

\begin{equation}
    A=\left[\frac{\Gamma(n)}{\Gamma(n+1-\gamma)}\right]^2 n^{2(1-\gamma)}.
    \label{A}
\end{equation}

\noindent Therefore, the equation for the Hubble parameter is\footnote{
Equation~\eqref{H_final} can of course be rewritten in such a way so that 
\( \Omega_\text{M}^\star = 1 \), where \( \Omega_\text{M}^\star \) is the 
right-hand side of equation~\eqref{H_final} divided by \( H \), but as we
mentioned before we preferred to stay with the definitions used in 
standard cosmology.  Furthermore, the use of equation~\eqref{H_final} is  
very convenient since in it, the Hubble parameter \( H \) is  represented by
a simple  function of \( \Omega_\text{M} \).  A similar relation is not found in 
standard cosmology and a value for the Hubble constant needs to be known
somehow  when fitting the standard cosmological model to SN~Ia data as 
shown in the Appendix~\ref{CDM}}:

\begin{equation}
    H=\frac{a_0}{c}\left(\Omega_\text{M}\right)^{1/2(\gamma-1)}.
    \label{H_final}
\end{equation}

\noindent In the previous equation, the definitions of the density parameters given 
in~\eqref{densities}  were used.  Nonetheless it is possible to define 
suitable new adequate density parameters in order to obtain an equation
similar to~\eqref{suma1}, as mentioned on Footnote~\ref{foot01}. 
Also, the use of equuation~\eqref{H_final} is quite useful since 
the definitions of the fractional 
cosmographic parameters in equations~\eqref{frac_q}, \eqref{fraction_j} 
and~\eqref{fraction_s},  involve the Hubble parameter $H$, making necessary 
an explicit equation for such parameter. Note that the previous 
equation is not valid for \( \gamma = 1 \) and since we are going to use that 
result on many of the rest of the article, in here and in 
what follows we demand \( \gamma \neq 1 \).

  By defining a fractional deceleration parameter as:

\begin{equation}
    q^\star := -\frac{1}{aH^{2\gamma}}\frac{D^\gamma}{Dt^\gamma}
      \left(\frac{D^\gamma a}{Dt^\gamma}\right), 
\label{frac_q}
\end{equation}

\noindent where $\gamma$ in $H^{2\gamma}$ is an exponent, 
the second  fractional Friedmann equation~\eqref{fraction2} 
for a matter dominated Universe can be 
written as:

\begin{equation}
  q^\star H^{2\gamma}=\kappa\frac{4\pi G \rho}{3},
\label{fried_q_frac}
\end{equation}

\noindent which after dividing by \( H^2 \) yields: 

\begin{equation}
  q^\star H^{2(\gamma-1)}=\kappa\frac{1}{2}
    \Omega_\text{M},
\label{fried_q_frac2}
\end{equation}

\noindent and so, using equation~\eqref{H_final} and \eqref{kappa} we find:

\begin{equation}
    q^\star = \frac{A}{2}.
\label{q_density}
\end{equation}

  In order to find expressions for the fractional jerk and snap 
cosmographic parameters as functions of the density parameters only,
the natural way would be to follow an analogous procedure as the one 
described in Section~\ref{general-relativity}. This procedure will involve
the cumbersome application of two consecutive fractional cosmic derivatives 
(one for the jerk and another for the snap)
in equation~\eqref{fraction2}. 
To avoid that, we apply a Last Step Modification procedure in 
equations~\eqref{derivative1} and~\eqref{derivative2} with correct 
definitions for the  fractional jerk and snap parameters. For 
simplicity and coherence, we will continue to use in what follows a 
Last Step Modification and so,  the fractional equivalent of relation~\eqref{derivative1} is given by:

\begin{eqnarray}
    \frac{1}{a}\frac{D^\gamma}{Dt^\gamma}\left(\frac{D^\gamma}{Dt^\gamma}\left(\frac{D^\gamma a}{Dt^\gamma}\right)\right)
    -\frac{1}{a^2}\frac{D^\gamma}{Dt^\gamma}\left(\frac{D^\gamma a}{Dt^\gamma}\right)\frac{D^\gamma a}{Dt^\gamma} \nonumber \\
    = -\kappa \frac{4\pi G }{3}\frac{D^\gamma \rho}{Dt^\gamma}.
\label{frac_deriv}
\end{eqnarray}

  We define the fractional jerk parameter as:

\begin{equation}
    j^\star := \frac{1}{aH^{3\gamma}}\frac{D^\gamma}{Dt^\gamma}\left(\frac{D^\gamma}{Dt^\gamma}\left(\frac{D^\gamma a}{Dt^\gamma}\right)\right),
\label{fraction_j}
\end{equation}

\noindent and substitute this together with  the definitions of the fractional deceleration~\eqref{frac_q}
and Hubble parameter~\eqref{fracH} into equation~\eqref{frac_deriv} to obtain:

\begin{equation}
    H^{3\gamma}j^\star + q^\star H^{2\gamma}H^\star= -\kappa \frac{4\pi G }{3}\frac{D^\gamma \rho}{Dt^\gamma}.
    \label{aux1.1}
\end{equation}

  In order to obtain an expression for $D^\gamma \rho/Dt^\gamma$
an equation for $\rho$ must be found. From equation \eqref{H_final} the following 
relation is given:

\begin{equation}
    \rho=\frac{3}{8\pi G}\left(\frac{c}{a_0}\right)^{2(\gamma-1)}H^{2\gamma}.
    \label{rho}
\end{equation}

  With the ansatz $a=a_1t^n$, the fractional derivative of order $\gamma$
of the previous equation is:

\begin{equation}
    \frac{D^\gamma \rho}{Dt^\rho}=\frac{3}{8\pi G}\left(\frac{c}{a_0}\right)^{2(\gamma-1)}
    n^{2\gamma}\frac{\Gamma(1-2\gamma)}{\Gamma(1-3\gamma)}t^{-3\gamma},
    \label{rho_gamma}
\end{equation}

\noindent or in terms of the Hubble parameter and the matter density:

\begin{equation}
   \frac{D^\gamma \rho}{Dt^\gamma}=\frac{\Gamma(1-2\gamma)}{n^\gamma\Gamma(1-3\gamma)}H^\gamma\rho.
   \label{Drho}
\end{equation}

\noindent With this result, equation~\eqref{aux1.1} takes the following form:

\begin{equation}
    H^{2\gamma}j^\star + q^\star H^\gamma H^\star= 
    -A \left(\frac{a_0}{c}\right)^{2(\gamma-1)} \frac{\Gamma(1-2\gamma)}{n^\gamma\Gamma(1-3\gamma)}\frac{4 \pi G}{3}\rho.
    \label{aux1}
\end{equation}

\noindent or: 

\begin{eqnarray}
    H^{2(\gamma-1)}\left[j^\star + q^\star \frac{\Gamma(n)}{\Gamma(n+1-\gamma)n^{\gamma-1}}\right] &=& \nonumber \\
    -A \left(\frac{a_0}{c}\right)^{2(\gamma-1)}\frac{\Gamma(1-2\gamma)}{n^\gamma\Gamma(1-3\gamma)}\frac{\Omega_{\text{M}}}{2}.
    \label{aux2}
\end{eqnarray}

\noindent where equation \eqref{intermedio2} was used. Direct substitution of 
equations \eqref{H_final} and \eqref{q_density} into the previous relation yields:

\begin{equation}
   j^\star=-\frac{A}{2 n^\gamma}\left[ \frac{\Gamma(1-2\gamma)}{\Gamma(1-3\gamma)}+
   \frac{\Gamma(n)}{\Gamma(n+1-\gamma)n^{-1}} \right].
\label{fraction_jone}
\end{equation}

% \noindent That is, the value of the jerk parameter is not influenced by the fractional derivatives with respect
% to its value given by general relativity.

  A fractional analogous of equation \eqref{derivative2} involves the term $D^\gamma_t D^\gamma_t \rho$.
Since the Leibniz's rule is a complicated relation (see Appendix~\ref{apen1}), it is easier to derive 
equation \eqref{rho_gamma} instead of \eqref{Drho}.

  The fractional snap parameter is defined by:

\begin{equation}
    s^\star := \frac{1}{aH^{4\gamma}}\frac{D^\gamma}{Dt^\gamma}\left(\frac{D^\gamma}{Dt^\gamma}
    \left(\frac{D^\gamma}{Dt^\gamma}\left(\frac{D^\gamma a}{Dt^\gamma}\right)\right)\right).
\label{fraction_ss}
\end{equation}

\noindent By performing a similar procedure  as the one to obtain the fractional jerk parameter it follows that:

\begin{eqnarray}
    s^\star&=& -A\left[\frac{\Gamma(1-2\gamma)}{2\Gamma(1-4\gamma)n^{2\gamma}}\right. \nonumber\\
    & &\left. + \frac{\Gamma(n)\Gamma(1-2\gamma)}{\Gamma(1-3\gamma)\Gamma(n+1-\gamma)n^{2\gamma-1}}
    -\frac{A}{2}\right].
\label{fraction_s} 
\end{eqnarray}

\section{SN~Ia fits}
\label{cosmography}

  The accelerated expansion of the Universe was first inferred through
cosmological observations of SN~Ia as standard candles at all 
redshifts. \citet{Perlmutter99} used 42 high-redshift supernovae to construct
an apparent magnitude-redshift diagram 
in order to obtain the best values for the matter density parameter $\Omega_\text{M}$
with the introduction of a  dark energy density parameter $\Omega_\Lambda$ component.

  The model independent relation between the luminosity distance 
\( d_\text{L}(z) \) as a function of redshift \( z  \) and the apparent magnitude 
\( \mu \) is given by \citep{Peebles,Barrientos20}:

\begin{equation}
    \mu(z) = 5 \log \left[\frac{H_0 d_L(z)}{c}\right] - 5 \log h(z) + 42.3856,
    \label{modulus}
\end{equation}

\noindent where $H_0$ is the Hubble constant evaluated at the present epoch 
\( t_0 \) and $h := H_0 / 100 \textrm{km/s} / \mathrm{Mpc} $ is the normalised Hubble constant, with the luminosity distance for a flat universe given by~\citep{Visser}:

\begin{equation}
  \begin{split}
    d_L(z) = \frac{c}{H_0} \left[z+\frac{1}{2}(1-q_0)z^2
    -\frac{1}{6} \left(1-q_0-3q_0^2+j_0 \right)z^3\right. 
    \\
    \left. +\frac{1}{24} \left(2-2q_0-15q_0^2-15q_0^3+5j_0+10q_0 j_0+s_0\right)z^4 + ...\right], 
  \end{split}
\label{distance}
\end{equation}

\noindent where $q_0$, $j_0$ and $s_0$ are the cosmographic parameters evaluated at the present epoch. 
Substitution of  equations~\eqref{deceleration}, \eqref{jerk} and~\eqref{snap} into the previous relation yields 
an expression for the distance modulus $\mu$ as a function of 
$z$ and $\Omega_\text{M}$. Since the values for the 
distance modulus and redshift are given by observations, a statistical fit will return the 
best estimated values for both density parameters. 

 Since our intention is to see whether the fractional Friedmann model presented above 
can account for SN~Ia observations, we made the assumption that the Last
Step Approximation can be employed on equation~\eqref{distance} and the
cosmographic parameters defined in terms of the ordinary time derivatives
can be replaced by their fractional analogues, so instead of using equations~\eqref{deceleration}, 
\eqref{jerk} and \eqref{snap} for the cosmographic deceleration, jerk and snap parameters, 
we use their fractional extension given in~\eqref{q_density}, \eqref{fraction_jone}  
and~ \eqref{fraction_s}. This generalisation requires 
the Hubble parameter $H_0$ to be  given  by equation~\eqref{H_final}.
The SN~Ia distance modulus-redshift data 
was taken from the Supernova Cosmology Project (SCP) 
Union 2.1~\citep{SCP-Union2012}\footnote{ The data can be
downloaded  at the following  location: \url{http://supernova.lbl.gov/Union/figures/SCPUnion2.1_mu_vs_z.txt},
or in the cosmology 
tables section of \url{http://supernova.lbl.gov/Union/} }.

  In order to perform the statistical fits, we used the free software
\textit{gnuplot} (\url{www.gnuplot.info}) to obtain the best 
values for the free parameters of our model.  The calibration can be 
directly performed since the constants $c$, $a_0$ and $j_0$ are known and so 
the corresponding equations for $H_0$, $q_0$ and $s_0$ become 
functions of the density parameters and the fractional order. 
We used the  fit function command 
in \textit{gnuplot} for the calibration of the free parameters. 
This command uses non-linear
and linear least squares methods and is able to fit the required function
through the empirical data and provides the correlation 
matrix between the parameters, the number of iterations employed for the converged fit, the final sum 
of the squares of residuals (SSR), the best fit value for the parameters and its asymptotic standard error,  
the $p$-value for the $\chi$-square distribution and the root mean squares of residuals.

  At this point it is very important to note that in the definition of the cosmographic
parameters \( q^\star \), \( j^\star \) and \( s^\star \) we used the Hubble parameter 
\( H \) and not \( H^\star \).  With the use of the Last Step Modification technique 
we have no formal way to decide which one to use.  We have used \( H \) for the simplicity
it produces but of course at first sight it seems more natural that the choice 
\( H^\star \) would be the correct choice.

  In order to see that the choice \( H \) is the only viable one we proceed as follows.
The final relation we would like to solve is the redshift-magnitude relation~\eqref{modulus}.  
That equation contains the Hubble parameter \( H(z) \) through the definition of \( h \).  
As mentioned on Appendix~\ref{CDM}, for the case of the standard \( \Lambda \)-CDM cosmology 
a knowledge of \( H_0 \) is required to perform the fit to SN~Ia data since there is no 
independent equation to relate \( H(z) \) with some of the parameters of the model.  In the 
case of fractional calculus cosmology studied in this article we could have chosen to use 
\( H^\star \) instead of \( H \) on equation~\eqref{modulus}, but then we would be required
to know by observations the value of \( H_0^\star \).  In order to avoid this it is then 
best to leave equation~\eqref{modulus} as it is and take advantage of relation~\eqref{H_final}, 
i.e. in our model we do not need to know a priori the value of \( H_0 \) to perform the fit
to SN~Ia data.

  Also, it is important to note that the definitions of the csomographic parameters 
\( q^\star \), \( j^\star \)  and \( s^\star \) in 
equations~\eqref{frac_q},~\eqref{fraction_j} and~\eqref{fraction_ss} 
could have been made in terms of \( H^\star \) instead of \( H \).
But this just means to rewrite 
the standard cosmological equations~\eqref{suma1},~\eqref{deceleration},~\eqref{jerk} 
and~\eqref{snap} with their star counterparts. This will not have any departure 
from the standard $\Lambda$CDM
cosmological model. 

\section{Results}
\label{sub_resul}

% \subsection{Free order fractional derivative.}

The fractional cosmographic parameters are functions 
that depend on $n$ (via $A$), $\gamma$ and \( \Omega_M \). 
Therefore, an extensive search for a coherent 
fit with the simple \textit{gnuplot} routine with those 
three parameters as free variables was made. Setting up the initial 
values as follows: $n=2.6$, $\gamma=2.1$ and $\Omega_\text{M}=4.5$, 
yields quite good convergence for the routine. The obtained results 
are reported in Table~\ref{2_parameters}.

%In an extensive
%search for a coherent fit with the simple \textit{gnuplot} routine we found that with
%a value of \( n \) between \( 0.6  \)--\( 10.0 \), the  best results are 
%obtained and are very similar to one another.  Due to the non-linearity of the 
%equations and since our model is a first approach toy model,  we decided to 
%fix \( n \) using the following arguments.  If we assume that the mass 
%conservation equation of standard cosmology is valid, then \( \rho \propto
%a^{-3} \) and since \( a \propto t^n \), then equation~\eqref{fraction1} 
%yields \( n = 2 \gamma / 3 \).  By doing this, the obtained values are:
%\textcolor{red}{\( \gamma = 1.6771 \pm 0.1893 \)} and \textcolor{red}{\( \Omega_M = 19.9823 \pm 16.69 \)}, with
%an SSR of \textcolor{red}{\( 702.507 \)} and a $p$-value of \textcolor{red}{\( 0.0002 \)}.  The large error 
%obtained in \( \Omega_M \)  is most probably due to the fact that \( n \neq 
%2 \gamma /3 \), i.e. the mass conservation equation requires to be expressed 
%in fractional derivatives.  To avoid the complication that this will imply, we
%left \( n \) and \( \gamma \) as free parameters of the model and made an 
%extensive search about the value of \( \gamma \sim 3/2 \) which by \textcolor{red}{assuming}
%the standard mass conservation equation of cosmology yields \( n \sim 1.0 \).
%It turns out that we obtained quite a good convergence by choosing
%\( n = 1.1 \),
%providing starting given values to 
%\textit{gnuplot} around \( \Omega_\text{M} = 9.5 \) and \( \gamma = 2.5 \).
%The obtained results are reported in Table~\ref{2_parameters}.

\begin{figure}
  \includegraphics[scale=0.7]{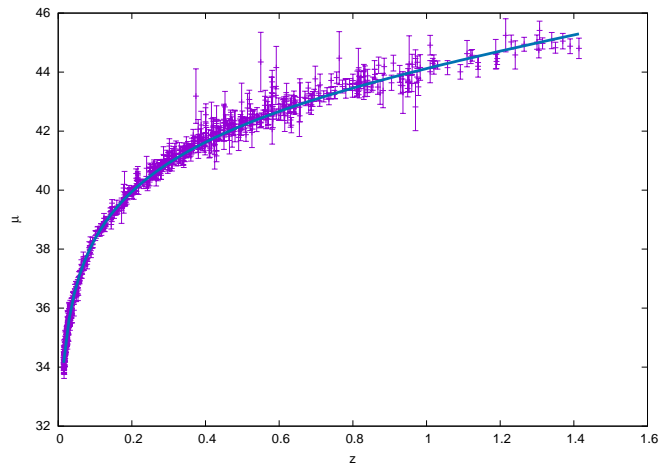}
  \caption[Figure01]{Apparent magnitude \( \mu \) vs.~redshift \(
z \) Hubble diagram from the Union 2.1 SNe~Ia data (dots with their
corresponding error bars) and the best fit from our model. The solid
line represents the distance modulus $\mu(z)$ from the best fit to the
data of the model.}
	\label{fig_gamma_free}
\end{figure}

Although the error in \( n \) is \( \sim 90\% \), the mean plotted
best distance modulus-redshift function in Figure~\ref{fig_gamma_free} 
seems to be in a good agreement with the observations.

\begin{table}
	\begin{center}
		\begin{tabular}{|c|c|}
		\hline
		$\gamma$ & $1.7254\pm 0.043$\\
		\hline
		$\Omega_\text{M}$  & $11.6273\pm 1.583$ \\
		\hline
		$n$ & $0.9306 \pm 0.846$\\
		\hline
		\end{tabular}
		\hspace{1.0cm}
        \begin{tabular}{|c|c|c|c|}		
		 \hline
         & $\gamma$ &  $\Omega_\text{M}$ &n\\ 
         \hline
         $\gamma$ & 1.000 & &\\
         $\Omega_\text{M}$ & 0.999 & 1.000 &\\
         $n$ & 0.995 & 0.995 & 1.000\\
         \hline
		\end{tabular}
	\end{center}
\caption[Table01]{Left: Best fit results for the fractional derivative model with three 
free parameters: the fractional  derivative order \( \gamma \), the
matter density parameter \(\Omega_\text{M} \) and the power for the scale factor $n$
presented with their corresponding errors for the initial values: $n=2.6$, $\gamma=2.1$ and $\Omega_\text{M}=4.5$. 
The right	panel shows the correlation matrix for the best 
fit values reported. The SSR and the $p$-value for this model are: 
\(636.015 \) and  $0.04$ 
respectively}. 
\label{2_parameters}
\end{table}

  The envelope curves that represent the statistical errors above and below the
solid mean curve in Figure~\ref{fig_gamma_free} have not been drawn since
they turn out to be quite wide due to the somewhat large error in \( n\).

With the mean values reported in Table~\ref{2_parameters},
the Hubble constant $H_0$ given by equation \eqref{H_final} has the following numerical value:

\begin{equation}
    H_0=66.95 \,\mbox{km/s/Mpc},
    \label{H0result}
\end{equation}

\noindent which is in a great agreement with the value reported by Planck \citep{planck} 
(see Appendix~\ref{CDM}).

\begin{figure}
  \includegraphics[scale=0.7]{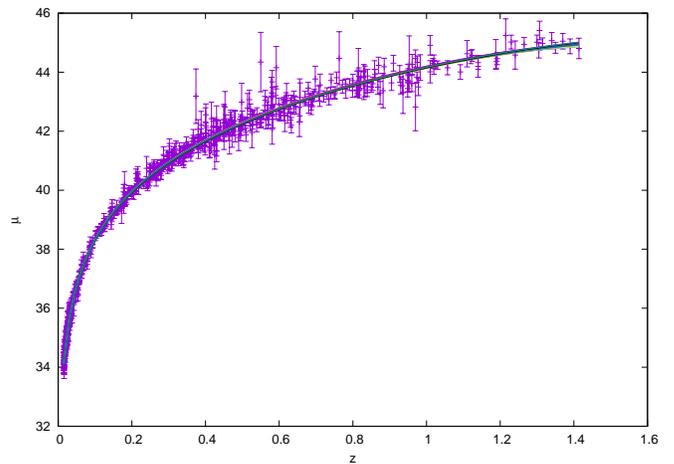}
  \caption[Figure02]{Apparent magnitude \( \mu \) vs.~redshift \(
z \) Hubble diagram from the Union 2.1 SNe~Ia data (dots with their
corresponding error bars) and the best fit from our model. The solid
blue line represents the distance modulus $\mu(z)$ from the best fit to the
data of the model. The other solid curves represents the envelope of the statistical
errors.}
	\label{errors}
\end{figure}

Due to limitations of the \textit{gnuplot}'s fit routine and the divergences in the
range $x<0$ of the $\Gamma(x)$ function, the convergence of the fit is highly dependent
on the initial values of the free parameters $n$, $\gamma$ and $\Omega_\text{M}$. 
The results shown in Table~\ref{2_parameters} are obtained using a wide range of initial 
values. Another fit with a good convergence of the fit is given by the initial values:
$n=1.3$, $\gamma=2.6$ and $\Omega_\text{M}=2.1$. The obtained values are reported
in Table \ref{table_error}. The distance modulus-redshift plot for the values 
obtained in this fit is shown in Figure \ref{errors}. The envelope curves that represent the 
statistical errors are present, but due to the small asymptotic errors such envelope 
is cover by the mean curve.

\begin{table}
	\begin{center}
		\begin{tabular}{|c|c|}
		\hline
		$\gamma$ & $1.4937\pm 0.0003$\\
		\hline
		$\Omega_\text{M}$  & $5.4220\pm 0.0243$ \\
		\hline
		$n$ & $0.5539 \pm 0.0046$\\
		\hline
		\end{tabular}
		\hspace{1.0cm}
        \begin{tabular}{|c|c|c|c|}		
		 \hline
         & $\gamma$ &  $\Omega_\text{M}$ &n\\ 
         \hline
         $\gamma$ & 1.000 & &\\
         $\Omega_\text{M}$ & 0.273 & 1.000 &\\
         $n$ & -0.140 & 0.587 & 1.000\\
         \hline
		\end{tabular}
	\end{center}
\caption[Table02]{Left: Best fit results for the fractional derivative model with three 
free parameters: the fractional  derivative order \( \gamma \), the
matter density parameter \(\Omega_\text{M} \) and the power for the scale factor $n$
presented with their corresponding errors for the initial values: $n=1.3$, $\gamma=2.6$ and $\Omega_\text{M}=2.1$. 
The right	panel shows the correlation matrix for the best 
fit values reported. The SSR and the $p$-value for this model are:  \(573.544 \) and  $0.50$ 
respectively}. 
\label{table_error}
\end{table}

For this set of initial values the resulting asymptotic error of the free parameters 
is considerably less than those reported in Table\ \ref{2_parameters}. The parameters $\gamma$ and 
$\Omega_\text{M}$  do not have a significant improvement, unlike the parameter \( n \) whose error
drastically decreased. Another relevant difference resides in the p-value, this statistical
indicator increased, but is acceptable to within the ranges of \textit{gnuplot}. With these results, the Hubble constant has the following value:

\begin{equation}
    H_0=68.37 \,\mbox{km/s/Mpc}.
    \label{H0result2}
\end{equation}

\section{Final remarks}
\label{discussion}

  In this article we have generalised the standard Friedmann equations of 
cosmology using fractional derivatives by performing a Last Step 
Modification approach.   The obtained fractional Friedmann equations 
and the formulae 
of the corresponding fractional cosmographic parameters in terms of the
cosmological density parameters were used to explain the current accelerated
expansion of a dust and zero curvature Universe, without the 
introduction of a dark energy component.  We introduced Milgrom's acceleration 
constant into the built expressions with the intention of not requiring any non-baryonic dark 
matter component.
The statistical 
fitting of the free parameters in the model shows an excellent agreement with a fractional
derivative order of \( 3/2 \), which has recently been shown to be the 
required fractional order to explain MOdified Newtonian Dynamics (MOND) 
phenomenology~\citep{andrea}.    This result is in excellent agreements with the recent work
by~\citet{Barrientos20} since the Universe at the present epoch is in the 
deep MOND regime\footnote{ 
  \citet{Milgrom1} noticed a coincidental relation between the acceleration
\( a_0 \) and \( H_0 \): \( a_0 \approx c H_0 / 2 \pi \).  Since the Hubble
radius \( r_H = c / H_0 \) and the Hubble mass is \( M_H \approx c^3 / 2 G H_0 \)
it then follows that the Newtonian gravitational acceleration experienced
by a test galaxy at a distance  \( r_H \) of a point mass source \( M_H \)
is \( \approx G M_H / r_H^2 \approx \pi a_0 \) \citep[see
e.g.][]{Milgrom:2008,Bernal:2011a}.  In other words, the Newtonian
gravitational acceleration at the present epoch in the Universe is
approximately MONDian.  As such, one would expect that if no dark matter
components are introduced into the cosmological energy budget, then a
MONDian description of the Universe at the present epoch is required.
}. The introduction of fractional derivatives into physics
usually means that non-locality is taking place at some level and it gets 
more noticeable at large scales.  As noted by \citet{Barrientos20}, there are
an infinite number of non-local relativistic theories of gravity with
curvature-matter couplings that have MOND on their weak field limit of 
approximation.  All this is pointing into the direction that MOND is 
most probably a non-local phenomenon which occurs at sufficiently small mass
to squared length scales, i.e. at acceleration scales \( \lesssim a_0 \).

  Strictly speaking, the fitting procedure is only giving information about the total
matter density \( \Omega_\textrm{M} \) which includes baryonic and non-baryonic components.  
In this sense the model presented shows that a simple introduction of fractional Friedmann 
equations in cosmology can account for a dark energy component.  However, as mentioned in 
Section~\ref{fractional-friedmann}, the introduction of Milgrom's acceleration constant 
\( a_0 \) in equation~\eqref{kappa} was done with the intention that 
the dark matter component would be taken into account by a MOND-like prescription.  The obtained
value of the fractional derivative order of \( \sim 3/2 \) in this article  and the fact that
we expect MONDian effects to be relevant in the present epoch dynamics of the Universe, 
further reinforce this result as mentioned in the previous paragraph.

    The present work is restricted to a matter dominated Universe and 
puts aside the possibility of a dark-energy dominated Universe. The main reason lies in the
complexity of the resultant equations.  In fact, for a \(\Lambda\)-dominated Universe, 
equation~\eqref{ansatz} is not longer valid. 
Thinking in analogy to the standard cosmological model, we can expect that the
scale factor for a dark-energy dominated era has an exponential dependence on $t$. 
However, the fractional derivative of an exponential function is a cumbersome expression that makes 
the fit extremely difficult. The case of a Universe where both components are taken into account
is equally complex since our definitions for the cosmographic parameters require an equation 
for the Hubble parameter and equation~\eqref{intermedio1} is a relation for $H^\star$.
In order to have a more solid proposal, a more profound model needs to be built, introducing 
fractional derivatives in the principle of least action. 
Such topic goes beyond to the toy model introduced in this  work.

  The toy model presented in this article about the current expansion of the Universe using 
fractional derivatives using the Last Step Modification is to be taken with
care.  It represents a first exploration onto whether there 
could be any interesting aspects of gravitational theory to be more deeply 
investigated since the field equations do not come from a variational principle.  
We intend to go beyond 
the present work to construct a formalism in terms of fractional derivatives for 
geometrical objects in general relativity and in formulating a fractional 
calculus of variations with applications to gravitational theory.

\section*{Acknowledgements}
The authors acknowlodge the excellent comments made by three reviewers for 
useful suggestions that  improved the final version of the article.  
This work was supported
by DGAPA-UNAM (IN112019) and CONACyT (CB-2014-01 No.~240512) 
grants. EB thanks support from a Postdoctoral fellowship granted by CONACyT.  
EB, SM and PP acknowledge economic support from 
CONACyT (517586, 26344, 14558).

%%%%%%%%%%%%%%%%
% BIBLIOGRAPHY %
%%%%%%%%%%%%%%%%
% \bibliographystyle{apsrev4-1}
%\bibliographystyle{aipauth4-1}
\bibliographystyle{aipnum4-1}
\bibliography{fractional-cosmology}

\appendix

\section{Fractional calculus: a simple introduction.}
\label{apen1}

The question of whether it is possible to take half the derivative of a function goes back to a letter written by Leibniz 
to L'Hôpital in 1665. Since then the subject of fractional calculus, as it is now known, has attracted the interest of mathematicians like Abel, Liouville and Riemann, who developed the basis of
the theory. In recent years, fractional derivatives have proved to be useful in many applications ranging from anomalous diffusion, finance to modeling of viscoelastic materials. As expected,when considering the order of differentiation to be an integer, the fractional derivative coincides with the standard definition. However, for non-integer values it has an  intrinsic non-local character, which explains its applicability in complex phenomena in which long range interactions or memory effects are present.
Before providing the general definition of fractional derivative, we give a few examples motivating this notion.
Take for instance 
$$f(x)=x^k.$$
Computing the first derivative we obtain
$$f^\prime(x)= \frac{df}{dx}= k x^{k-1}.$$
The successive repetition of  this procedure $n$ times yields the $n$-th derivative:
$$ 
\frac{d^n f}{dx^n}= \frac{k!}{(k-n)!} x^{k-n}.
$$
The above expression makes perfect sense for a real number $k$, if we can define its
factorial in a meaningful way. The $\Gamma$ function
provides such a generalization, and the above expression can be written as

$$ 
\frac{d^\alpha f}{dx^\alpha}= \frac{\Gamma(k+1)}{\Gamma(k-\alpha+1)} x^{k-\alpha},
$$
where we have replaced $n$ by $\alpha$ to emphasize the fact that now the order of differentiation can be taken to be any real number.\\
A similar reasoning can be applied to other basic examples such as the exponential or trigonometric functions. In such a way, it is possible to define the fractional derivative or integral for functions expressed in their Taylor or Fourier series expansions. However, a more general and useful definition is based on the Riemann-Liouville representation formula of a function.\ More precisely, the integral operator of order $\alpha$ is defined as
$$
I^\alpha f(t)= \frac{1}{\Gamma(\alpha)} \int_0^t \frac{f(\tau)}{(t-\tau)^{1-\alpha}}\, d\tau.
$$
Again, this definition can be justified by applying the standard integral operator $m$ times and integrating by parts and then making sense of the definition for a real order of integration.
Less intuitive, but more appropriate in formulating initial value problems than other definitions of fractional derivatives is the Caputo proposal given by
$$
D^\alpha f(t)= I^{m-\alpha}D^m f= \frac{1}{\Gamma(m-\alpha)} \int_0^t (t-\tau)^{m-\alpha-1} f^{(m)}(\tau)\, d\tau, 
$$
for $m-1<\alpha<m$, $m\in \mathbb{N}$ and where $f^{(m)}$ denotes the standard derivative of $f$. 
For a systematic presentation of fractional calculus the reader is referred to~\citet{Podlubny}.

There are two properties of fractional derivatives that are important to mention since 
they depar from the common sense in ordinary calculus . The first one is 
the Leibniz's rule for the product of two functions. \citet{oldham} proved that the most general expression is given by:

\begin{equation*}
    \frac{D^\alpha[fg]}{dx^\alpha}=\sum_{j=-\infty}^\infty\frac{\Gamma(\alpha+1)}{\Gamma(\alpha-\beta-j+1)\Gamma(\beta+j+1)}
    \frac{d^{\alpha-\beta-j}f}{dx^{\alpha-\beta-j}}\frac{d^{\beta+j}g}{dx^{\beta+j}}.
    \label{leib}
\end{equation*}

The second one is the composition rule:

\begin{equation*}
    D^\alpha D^\beta f=D^{\alpha+\beta}f-\sum_{k=0}^{\beta-1}\frac{[x-a]^{k-\alpha-\beta} f^{(k)}(a)}{\Gamma(k-\alpha-\beta+1)}.
    \label{compo}
\end{equation*}

\section{$\Lambda CDM$ standard cosmology.}
\label{CDM}

With the intention of showing the robustness of the fitting procedure we are using
in the article with the fit routine of \textit{gnuplot}, we show in this appendix that for the case $\gamma=1$ we recover the values of $\Omega_M$ and 
$\Omega_\Lambda$ reported in the standard $\Lambda CDM$ cosmological model. To do so, we cannot simply fix 
the value $\gamma=1$ on all fractional equations since 
equation~\eqref{intermedio3} is not valid for such value. 
Therefore, the non-fractional equations~\eqref{suma1}, \eqref{deceleration}, \eqref{jerk} and~\eqref{snap} must be employed.
Unlike the fractional case, equation~\eqref{suma1} is a constriction and not an equation for the Hubble parameter 
$H_0$. Since the distance modulus $\mu(z)$ relation~\eqref{modulus} has an explicit dependence on this parameter, a value 
must be introduced. There are several values reported in the literature for $H_0$
that constitutes the so called ``\emph{cosmological tension}''. 
We  performed 
fits to SN~Ia data  using the values 
reported by the Planck Collaboration~\citep{planck}: $H_0^\text{Planck}=67.36\, \mbox{km s}^{-1} \mbox{Mpc}^{-1}$,
a value from SN~Ia observations: $H_0^\text{SN\,Ia}=63\, \mbox{km s}^{-1} \mbox{Mpc}^{-1}$ (see Fig. 9 in~\citep{Perlmutter99} and references
therein) and a local Hubble constant~\citep{camarena}: $H_0^\text{loc}=75.35\, \mbox{km s}^{-1} \mbox{Mpc}^{-1}$.

  Due to condition given by \eqref{suma1} the number of free parameters is reduced to one. 
$\Omega_M$ was chosen to be this free parameter. The results obtained by the fit are shown in Table \ref{fitCDM}. 

\begin{table}
	\begin{center}
        \begin{tabular}{|c|c|c|c|}		
	    \hline
		&Planck & SN\,Ia & Local \\
		\hline
		$\Omega_M$ & $0.185818$ & $ 0.3601 $ & $0.027163 $ \\
		\hline
		Asymptotic Error & $\pm 0.008958$ & $\pm 0.01343$ & $\pm 0.005725$ \\
		\hline
		SSR & $557.741$ & $822.58$ & $742.651$ \\
		\hline 
		p-value &  $0.203964$ & $<0.0001$ & $<0.0001$\\
		\hline
		\end{tabular}
	\end{center}
\caption[Table02]{ Best fit results for the standard cosmological model with one free parameter: the
matter density parameter \(\Omega_\text{M} \),  presented with its corresponding error, SSR and p-value.} 
\label{fitCDM}
\end{table}

The value of $\Omega_M^\text{loc}$ greatly differs from the standard estimation: $\Omega_M=0.27$.
However, such value is between $\Omega_M^{Planck}$ and $\Omega_M^{SN\,Ia}$. Thus, a Hubble parameter $H_0$ between
$63\, \mbox{km s}^{-1} \mbox{Mpc}^{-1}$ and $67.36\, \mbox{km s}^{-1} \mbox{Mpc}^{-1}$ will be closer to what is expected.
A recent study made by \citet{Mukherjee} reports the following estimations for the Hubble parameter and the matter density: 
$H_0=67.6^{+4.3}_{-4.2} \mbox{km s}^{-1} \mbox{Mpc}^{-1}$ and $\Omega_M=0.47^{+0.34}_{-0.27}$, which agree with the results
obtained from our fit.

\begin{figure}
    \centering
    \includegraphics[scale=0.7]{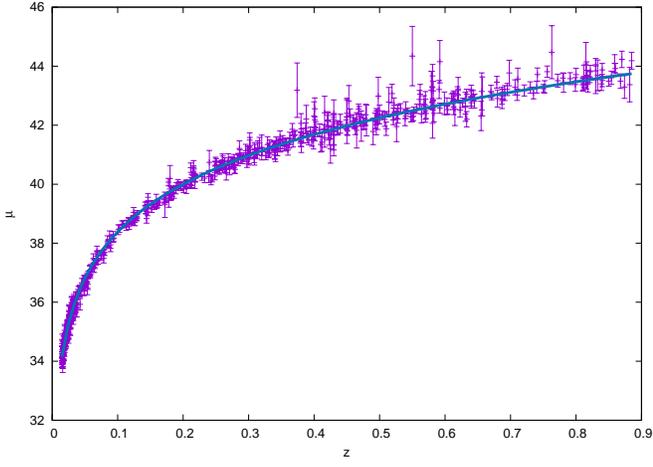}
    \caption[Figure:2]{Apparent magnitude \( \mu \) vs.~redshift \(
z \) Hubble diagram from the Union 2.1 SN~Ia data (dots with their
corresponding error bars) and the best fit from the $\Lambda CDM$ 
cosmological model with $H_0^\text{Planck}=67.36\, \mbox{km s}^{-1} \mbox{Mpc}^{-1}$. 
The solid line represents the distance modulus $\mu(z)$ 
from the best fit to the data.}
    \label{HubbleCDM}
\end{figure}

  Figure~\ref{HubbleCDM} shows the Hubble diagram for the $\Lambda CDM$ cosmological model using
the Planck value for the Hubble parameter.

\end{document}

%% file: aas-journals.tex
\newcommand{\apjl}{ApJL}

\newcommand{\mnras}{MNRAS}

\newcommand{\physrep}{Physics Reports}

%% file: fractional-cosmology.bbl
\begin{thebibliography}{82}%
\makeatletter
\providecommand \@ifxundefined [1]{%
 \@ifx{#1\undefined}
}%
\providecommand \@ifnum [1]{%
 \ifnum #1\expandafter \@firstoftwo
 \else \expandafter \@secondoftwo
 \fi
}%
\providecommand \@ifx [1]{%
 \ifx #1\expandafter \@firstoftwo
 \else \expandafter \@secondoftwo
 \fi
}%
\providecommand \natexlab [1]{#1}%
\providecommand \enquote  [1]{``#1''}%
\providecommand \bibnamefont  [1]{#1}%
\providecommand \bibfnamefont [1]{#1}%
\providecommand \citenamefont [1]{#1}%
\providecommand \href@noop [0]{\@secondoftwo}%
\providecommand \href [0]{\begingroup \@sanitize@url \@href}%
\providecommand \@href[1]{\@@startlink{#1}\@@href}%
\providecommand \@@href[1]{\endgroup#1\@@endlink}%
\providecommand \@sanitize@url [0]{\catcode `\\12\catcode `\$12\catcode
  `\&12\catcode `\#12\catcode `\^12\catcode `\_12\catcode `\%12\relax}%
\providecommand \@@startlink[1]{}%
\providecommand \@@endlink[0]{}%
\providecommand \url  [0]{\begingroup\@sanitize@url \@url }%
\providecommand \@url [1]{\endgroup\@href {#1}{\urlprefix }}%
\providecommand \urlprefix  [0]{URL }%
\providecommand \Eprint [0]{\href }%
\providecommand \doibase [0]{http://dx.doi.org/}%
\providecommand \selectlanguage [0]{\@gobble}%
\providecommand \bibinfo  [0]{\@secondoftwo}%
\providecommand \bibfield  [0]{\@secondoftwo}%
\providecommand \translation [1]{[#1]}%
\providecommand \BibitemOpen [0]{}%
\providecommand \bibitemStop [0]{}%
\providecommand \bibitemNoStop [0]{.\EOS\space}%
\providecommand \EOS [0]{\spacefactor3000\relax}%
\providecommand \BibitemShut  [1]{\csname bibitem#1\endcsname}%
\let\auto@bib@innerbib\@empty
%</preamble>
\bibitem [{\citenamefont {{Will}}(1993)}]{Will}%
  \BibitemOpen
  \bibfield  {author} {\bibinfo {author} {\bibfnamefont {C.~M.}\ \bibnamefont
  {{Will}}},\ }\href@noop {} {\emph {\bibinfo {title} {Theory and Experiment in
  Gravitational Physics, by Clifford M.~Will, pp.~396.~ISBN
  0521439736.~Cambridge, UK: Cambridge University Press, March 1993.}}}\
  (\bibinfo {year} {1993})\BibitemShut {NoStop}%
\bibitem [{\citenamefont {{Dyson}}, \citenamefont {{Eddington}},\ and\
  \citenamefont {{Davidson}}(1920)}]{Eddington}%
  \BibitemOpen
  \bibfield  {author} {\bibinfo {author} {\bibfnamefont {F.~W.}\ \bibnamefont
  {{Dyson}}}, \bibinfo {author} {\bibfnamefont {A.~S.}\ \bibnamefont
  {{Eddington}}}, \ and\ \bibinfo {author} {\bibfnamefont {C.}~\bibnamefont
  {{Davidson}}},\ }\href {\doibase 10.1098/rsta.1920.0009} {\bibfield
  {journal} {\bibinfo  {journal} {Philosophical Transactions of the Royal
  Society of London Series A}\ }\textbf {\bibinfo {volume} {220}},\ \bibinfo
  {pages} {291} (\bibinfo {year} {1920})}\BibitemShut {NoStop}%
\bibitem [{\citenamefont {{Pound}}\ and\ \citenamefont
  {{Rebka}}(1960)}]{Pound}%
  \BibitemOpen
  \bibfield  {author} {\bibinfo {author} {\bibfnamefont {R.~V.}\ \bibnamefont
  {{Pound}}}\ and\ \bibinfo {author} {\bibfnamefont {G.~A.}\ \bibnamefont
  {{Rebka}}},\ }\href {\doibase 10.1103/PhysRevLett.4.337} {\bibfield
  {journal} {\bibinfo  {journal} {\prl}\ }\textbf {\bibinfo {volume} {4}},\
  \bibinfo {pages} {337} (\bibinfo {year} {1960})}\BibitemShut {NoStop}%
\bibitem [{\citenamefont {{Reasenberg}}\ \emph {et~al.}(1979)\citenamefont
  {{Reasenberg}}, \citenamefont {{Shapiro}}, \citenamefont {{MacNeil}},
  \citenamefont {{Goldstein}}, \citenamefont {{Breidenthal}}, \citenamefont
  {{Brenkle}}, \citenamefont {{Cain}}, \citenamefont {{Kaufman}}, \citenamefont
  {{Komarek}},\ and\ \citenamefont {{Zygielbaum}}}]{Reasenberg}%
  \BibitemOpen
  \bibfield  {author} {\bibinfo {author} {\bibfnamefont {R.~D.}\ \bibnamefont
  {{Reasenberg}}}, \bibinfo {author} {\bibfnamefont {I.~I.}\ \bibnamefont
  {{Shapiro}}}, \bibinfo {author} {\bibfnamefont {P.~E.}\ \bibnamefont
  {{MacNeil}}}, \bibinfo {author} {\bibfnamefont {R.~B.}\ \bibnamefont
  {{Goldstein}}}, \bibinfo {author} {\bibfnamefont {J.~C.}\ \bibnamefont
  {{Breidenthal}}}, \bibinfo {author} {\bibfnamefont {J.~P.}\ \bibnamefont
  {{Brenkle}}}, \bibinfo {author} {\bibfnamefont {D.~L.}\ \bibnamefont
  {{Cain}}}, \bibinfo {author} {\bibfnamefont {T.~M.}\ \bibnamefont
  {{Kaufman}}}, \bibinfo {author} {\bibfnamefont {T.~A.}\ \bibnamefont
  {{Komarek}}}, \ and\ \bibinfo {author} {\bibfnamefont {A.~I.}\ \bibnamefont
  {{Zygielbaum}}},\ }\href {\doibase 10.1086/183144} {\bibfield  {journal}
  {\bibinfo  {journal} {\apjl}\ }\textbf {\bibinfo {volume} {234}},\ \bibinfo
  {pages} {L219} (\bibinfo {year} {1979})}\BibitemShut {NoStop}%
\bibitem [{\citenamefont {{Anderson}}\ \emph {et~al.}(1996)\citenamefont
  {{Anderson}}, \citenamefont {{Gross}}, \citenamefont {{Nordtvedt}},\ and\
  \citenamefont {{Turyshev}}}]{Anderson}%
  \BibitemOpen
  \bibfield  {author} {\bibinfo {author} {\bibfnamefont {J.~D.}\ \bibnamefont
  {{Anderson}}}, \bibinfo {author} {\bibfnamefont {M.}~\bibnamefont {{Gross}}},
  \bibinfo {author} {\bibfnamefont {K.~L.}\ \bibnamefont {{Nordtvedt}}}, \ and\
  \bibinfo {author} {\bibfnamefont {S.~G.}\ \bibnamefont {{Turyshev}}},\ }\href
  {\doibase 10.1086/176899} {\bibfield  {journal} {\bibinfo  {journal} {\apj}\
  }\textbf {\bibinfo {volume} {459}},\ \bibinfo {pages} {365} (\bibinfo {year}
  {1996})},\ \Eprint {http://arxiv.org/abs/gr-qc/9510029} {arXiv:gr-qc/9510029
  [gr-qc]} \BibitemShut {NoStop}%
\bibitem [{\citenamefont {Chandler}\ \emph {et~al.}(2005)\citenamefont
  {Chandler}, \citenamefont {Pearlman}, \citenamefont {Reasenberg},\ and\
  \citenamefont {Degnan}}]{Chandler}%
  \BibitemOpen
  \bibfield  {author} {\bibinfo {author} {\bibfnamefont {J.}~\bibnamefont
  {Chandler}}, \bibinfo {author} {\bibfnamefont {M.}~\bibnamefont {Pearlman}},
  \bibinfo {author} {\bibfnamefont {R.}~\bibnamefont {Reasenberg}}, \ and\
  \bibinfo {author} {\bibfnamefont {J.}~\bibnamefont {Degnan}},\ }\href@noop {}
  {\  (\bibinfo {year} {2005})}\BibitemShut {NoStop}%
\bibitem [{\citenamefont {{Ciufolini}}\ \emph {et~al.}(1998)\citenamefont
  {{Ciufolini}}, \citenamefont {{Pavlis}}, \citenamefont {{Chieppa}},
  \citenamefont {{Fernandes-Vieira}},\ and\ \citenamefont
  {{Perez-Mercader}}}]{Ciufolini}%
  \BibitemOpen
  \bibfield  {author} {\bibinfo {author} {\bibfnamefont {I.}~\bibnamefont
  {{Ciufolini}}}, \bibinfo {author} {\bibfnamefont {E.}~\bibnamefont
  {{Pavlis}}}, \bibinfo {author} {\bibfnamefont {F.}~\bibnamefont {{Chieppa}}},
  \bibinfo {author} {\bibfnamefont {E.}~\bibnamefont {{Fernandes-Vieira}}}, \
  and\ \bibinfo {author} {\bibfnamefont {J.}~\bibnamefont {{Perez-Mercader}}},\
  }\href {\doibase 10.1126/science.279.5359.2100} {\bibfield  {journal}
  {\bibinfo  {journal} {Science}\ }\textbf {\bibinfo {volume} {279}},\ \bibinfo
  {pages} {2100} (\bibinfo {year} {1998})}\BibitemShut {NoStop}%
\bibitem [{\citenamefont {{de Sitter}}(1916)}]{Sitter}%
  \BibitemOpen
  \bibfield  {author} {\bibinfo {author} {\bibfnamefont {W.}~\bibnamefont {{de
  Sitter}}},\ }\href {\doibase 10.1093/mnras/77.2.155} {\bibfield  {journal}
  {\bibinfo  {journal} {\mnras}\ }\textbf {\bibinfo {volume} {77}},\ \bibinfo
  {pages} {155} (\bibinfo {year} {1916})}\BibitemShut {NoStop}%
\bibitem [{\citenamefont {{Eubanks}}\ \emph {et~al.}(1997)\citenamefont
  {{Eubanks}}, \citenamefont {{Matsakis}}, \citenamefont {{Martin}},
  \citenamefont {{Archinal}}, \citenamefont {{McCarthy}}, \citenamefont
  {{Klioner}}, \citenamefont {{Shapiro}},\ and\ \citenamefont
  {{Shapiro}}}]{Eubanks}%
  \BibitemOpen
  \bibfield  {author} {\bibinfo {author} {\bibfnamefont {T.~M.}\ \bibnamefont
  {{Eubanks}}}, \bibinfo {author} {\bibfnamefont {D.~N.}\ \bibnamefont
  {{Matsakis}}}, \bibinfo {author} {\bibfnamefont {J.~O.}\ \bibnamefont
  {{Martin}}}, \bibinfo {author} {\bibfnamefont {B.~A.}\ \bibnamefont
  {{Archinal}}}, \bibinfo {author} {\bibfnamefont {D.~D.}\ \bibnamefont
  {{McCarthy}}}, \bibinfo {author} {\bibfnamefont {S.~A.}\ \bibnamefont
  {{Klioner}}}, \bibinfo {author} {\bibfnamefont {S.}~\bibnamefont
  {{Shapiro}}}, \ and\ \bibinfo {author} {\bibfnamefont {I.~I.}\ \bibnamefont
  {{Shapiro}}},\ }in\ \href@noop {} {\emph {\bibinfo {booktitle} {APS April
  Meeting Abstracts}}},\ \bibinfo {series and number} {APS Meeting Abstracts}\
  (\bibinfo {year} {1997})\ p.\ \bibinfo {pages} {K11.05}\BibitemShut {NoStop}%
\bibitem [{\citenamefont {{Nordtvedt}}(1968)}]{Nordtvedt}%
  \BibitemOpen
  \bibfield  {author} {\bibinfo {author} {\bibfnamefont {K.}~\bibnamefont
  {{Nordtvedt}}},\ }\href {\doibase 10.1103/PhysRev.170.1186} {\bibfield
  {journal} {\bibinfo  {journal} {Physical Review}\ }\textbf {\bibinfo {volume}
  {170}},\ \bibinfo {pages} {1186} (\bibinfo {year} {1968})}\BibitemShut
  {NoStop}%
\bibitem [{\citenamefont {{Nordtvedt}}(1973)}]{Nordtvedt2}%
  \BibitemOpen
  \bibfield  {author} {\bibinfo {author} {\bibfnamefont {K.}~\bibnamefont
  {{Nordtvedt}}},\ }\href {\doibase 10.1103/PhysRevD.7.2347} {\bibfield
  {journal} {\bibinfo  {journal} {\prd}\ }\textbf {\bibinfo {volume} {7}},\
  \bibinfo {pages} {2347} (\bibinfo {year} {1973})}\BibitemShut {NoStop}%
\bibitem [{\citenamefont {{Nordtvedt}}\ and\ \citenamefont
  {{Will}}(1972)}]{Nordtvedt3}%
  \BibitemOpen
  \bibfield  {author} {\bibinfo {author} {\bibfnamefont {J.}~\bibnamefont
  {{Nordtvedt}}, \bibfnamefont {Kenneth}}\ and\ \bibinfo {author}
  {\bibfnamefont {C.~M.}\ \bibnamefont {{Will}}},\ }\href {\doibase
  10.1086/151755} {\bibfield  {journal} {\bibinfo  {journal} {\apj}\ }\textbf
  {\bibinfo {volume} {177}},\ \bibinfo {pages} {775} (\bibinfo {year}
  {1972})}\BibitemShut {NoStop}%
\bibitem [{\citenamefont {{Schiff}}(1960)}]{Schiff}%
  \BibitemOpen
  \bibfield  {author} {\bibinfo {author} {\bibfnamefont {L.~I.}\ \bibnamefont
  {{Schiff}}},\ }\href {\doibase 10.1073/pnas.46.6.871} {\bibfield  {journal}
  {\bibinfo  {journal} {Proceedings of the National Academy of Science}\
  }\textbf {\bibinfo {volume} {46}},\ \bibinfo {pages} {871} (\bibinfo {year}
  {1960})}\BibitemShut {NoStop}%
\bibitem [{\citenamefont {{Shapiro}}(1964)}]{Shapiro}%
  \BibitemOpen
  \bibfield  {author} {\bibinfo {author} {\bibfnamefont {I.~I.}\ \bibnamefont
  {{Shapiro}}},\ }\href {\doibase 10.1103/PhysRevLett.13.789} {\bibfield
  {journal} {\bibinfo  {journal} {\prl}\ }\textbf {\bibinfo {volume} {13}},\
  \bibinfo {pages} {789} (\bibinfo {year} {1964})}\BibitemShut {NoStop}%
\bibitem [{\citenamefont {{Mendoza}}(2015)}]{mendoza15}%
  \BibitemOpen
  \bibfield  {author} {\bibinfo {author} {\bibfnamefont {S.}~\bibnamefont
  {{Mendoza}}},\ }\href {\doibase 10.1139/cjp-2014-0208} {\bibfield  {journal}
  {\bibinfo  {journal} {Canadian Journal of Physics}\ }\textbf {\bibinfo
  {volume} {93}},\ \bibinfo {pages} {217} (\bibinfo {year} {2015})}\BibitemShut
  {NoStop}%
\bibitem [{\citenamefont {Starobinsky}(1979)}]{staro}%
  \BibitemOpen
  \bibfield  {author} {\bibinfo {author} {\bibfnamefont {A.}~\bibnamefont
  {Starobinsky}},\ }\href@noop {} {\bibfield  {journal} {\bibinfo  {journal}
  {JETP Lett.}\ }\textbf {\bibinfo {volume} {30}},\ \bibinfo {pages} {682}
  (\bibinfo {year} {1979})}\BibitemShut {NoStop}%
\bibitem [{\citenamefont {{Nojiri}}\ and\ \citenamefont
  {{Odintsov}}(2011)}]{Nojiri11}%
  \BibitemOpen
  \bibfield  {author} {\bibinfo {author} {\bibfnamefont {S.}~\bibnamefont
  {{Nojiri}}}\ and\ \bibinfo {author} {\bibfnamefont {S.~D.}\ \bibnamefont
  {{Odintsov}}},\ }\href {\doibase 10.1016/j.physrep.2011.04.001} {\bibfield
  {journal} {\bibinfo  {journal} {\physrep}\ }\textbf {\bibinfo {volume}
  {505}},\ \bibinfo {pages} {59} (\bibinfo {year} {2011})},\ \Eprint
  {http://arxiv.org/abs/1011.0544} {arXiv:1011.0544 [gr-qc]} \BibitemShut
  {NoStop}%
\bibitem [{\citenamefont {{Nojiri}}\ and\ \citenamefont
  {{Odintsov}}(2008)}]{Nojiri08}%
  \BibitemOpen
  \bibfield  {author} {\bibinfo {author} {\bibfnamefont {S.}~\bibnamefont
  {{Nojiri}}}\ and\ \bibinfo {author} {\bibfnamefont {S.~D.}\ \bibnamefont
  {{Odintsov}}},\ }\href {\doibase 10.1016/j.physletb.2007.12.001} {\bibfield
  {journal} {\bibinfo  {journal} {Physics Letters B}\ }\textbf {\bibinfo
  {volume} {659}},\ \bibinfo {pages} {821} (\bibinfo {year} {2008})},\ \Eprint
  {http://arxiv.org/abs/0708.0924} {arXiv:0708.0924 [hep-th]} \BibitemShut
  {NoStop}%
\bibitem [{\citenamefont {Shamir}\ and\ \citenamefont {Fayyaz}(2020)}]{shamir}%
  \BibitemOpen
  \bibfield  {author} {\bibinfo {author} {\bibfnamefont {M.~F.}\ \bibnamefont
  {Shamir}}\ and\ \bibinfo {author} {\bibfnamefont {I.}~\bibnamefont
  {Fayyaz}},\ }\href {\doibase 10.1134/S0040577920010109} {\bibfield  {journal}
  {\bibinfo  {journal} {Theoretical and Mathematical Physics}\ }\textbf
  {\bibinfo {volume} {202}},\ \bibinfo {pages} {112} (\bibinfo {year}
  {2020})}\BibitemShut {NoStop}%
\bibitem [{\citenamefont {{Odintsov}}\ and\ \citenamefont
  {{Oikonomou}}(2020)}]{Odin20}%
  \BibitemOpen
  \bibfield  {author} {\bibinfo {author} {\bibfnamefont {S.~D.}\ \bibnamefont
  {{Odintsov}}}\ and\ \bibinfo {author} {\bibfnamefont {V.~K.}\ \bibnamefont
  {{Oikonomou}}},\ }\href {\doibase 10.1016/j.physletb.2020.135576} {\bibfield
  {journal} {\bibinfo  {journal} {Physics Letters B}\ }\textbf {\bibinfo
  {volume} {807}},\ \bibinfo {eid} {135576} (\bibinfo {year} {2020})},\ \Eprint
  {http://arxiv.org/abs/2005.12804} {arXiv:2005.12804 [gr-qc]} \BibitemShut
  {NoStop}%
\bibitem [{\citenamefont {{Nojiri}}\ \emph {et~al.}(2020)\citenamefont
  {{Nojiri}}, \citenamefont {{Odintsov}}, \citenamefont {{Oikonomou}},\ and\
  \citenamefont {{Popov}}}]{Nor-Odin20}%
  \BibitemOpen
  \bibfield  {author} {\bibinfo {author} {\bibfnamefont {S.}~\bibnamefont
  {{Nojiri}}}, \bibinfo {author} {\bibfnamefont {S.~D.}\ \bibnamefont
  {{Odintsov}}}, \bibinfo {author} {\bibfnamefont {V.~K.}\ \bibnamefont
  {{Oikonomou}}}, \ and\ \bibinfo {author} {\bibfnamefont {A.~A.}\ \bibnamefont
  {{Popov}}},\ }\href {\doibase 10.1016/j.dark.2020.100514} {\bibfield
  {journal} {\bibinfo  {journal} {Physics of the Dark Universe}\ }\textbf
  {\bibinfo {volume} {28}},\ \bibinfo {eid} {100514} (\bibinfo {year}
  {2020})},\ \Eprint {http://arxiv.org/abs/2002.10402} {arXiv:2002.10402
  [gr-qc]} \BibitemShut {NoStop}%
\bibitem [{\citenamefont {{Nojiri}}\ and\ \citenamefont
  {{Odintsov}}(2004)}]{Nor-Odin-04}%
  \BibitemOpen
  \bibfield  {author} {\bibinfo {author} {\bibfnamefont {S.}~\bibnamefont
  {{Nojiri}}}\ and\ \bibinfo {author} {\bibfnamefont {S.~D.}\ \bibnamefont
  {{Odintsov}}},\ }\href {\doibase 10.1023/B:GERG.0000035950.40718.48}
  {\bibfield  {journal} {\bibinfo  {journal} {General Relativity and
  Gravitation}\ }\textbf {\bibinfo {volume} {36}},\ \bibinfo {pages} {1765}
  (\bibinfo {year} {2004})},\ \Eprint {http://arxiv.org/abs/hep-th/0308176}
  {arXiv:hep-th/0308176 [hep-th]} \BibitemShut {NoStop}%
\bibitem [{\citenamefont {{Nojiri}}\ and\ \citenamefont
  {{Odintsov}}(2003)}]{Nor-Odin-03}%
  \BibitemOpen
  \bibfield  {author} {\bibinfo {author} {\bibfnamefont {S.}~\bibnamefont
  {{Nojiri}}}\ and\ \bibinfo {author} {\bibfnamefont {S.~D.}\ \bibnamefont
  {{Odintsov}}},\ }\href {\doibase 10.1103/PhysRevD.68.123512} {\bibfield
  {journal} {\bibinfo  {journal} {\prd}\ }\textbf {\bibinfo {volume} {68}},\
  \bibinfo {eid} {123512} (\bibinfo {year} {2003})},\ \Eprint
  {http://arxiv.org/abs/hep-th/0307288} {arXiv:hep-th/0307288 [hep-th]}
  \BibitemShut {NoStop}%
\bibitem [{\citenamefont {{Harko}}\ and\ \citenamefont {{Lobo}}(2010)}]{frlm}%
  \BibitemOpen
  \bibfield  {author} {\bibinfo {author} {\bibfnamefont {T.}~\bibnamefont
  {{Harko}}}\ and\ \bibinfo {author} {\bibfnamefont {F.~S.~N.}\ \bibnamefont
  {{Lobo}}},\ }\href {\doibase 10.1140/epjc/s10052-010-1467-3} {\bibfield
  {journal} {\bibinfo  {journal} {European Physical Journal C}\ }\textbf
  {\bibinfo {volume} {70}},\ \bibinfo {pages} {373} (\bibinfo {year} {2010})},\
  \Eprint {http://arxiv.org/abs/1008.4193} {arXiv:1008.4193 [gr-qc]}
  \BibitemShut {NoStop}%
\bibitem [{\citenamefont {{Harko}}\ \emph {et~al.}(2011)\citenamefont
  {{Harko}}, \citenamefont {{Lobo}}, \citenamefont {{Nojiri}},\ and\
  \citenamefont {{Odintsov}}}]{Harko1}%
  \BibitemOpen
  \bibfield  {author} {\bibinfo {author} {\bibfnamefont {T.}~\bibnamefont
  {{Harko}}}, \bibinfo {author} {\bibfnamefont {F.~S.~N.}\ \bibnamefont
  {{Lobo}}}, \bibinfo {author} {\bibfnamefont {S.}~\bibnamefont {{Nojiri}}}, \
  and\ \bibinfo {author} {\bibfnamefont {S.~D.}\ \bibnamefont {{Odintsov}}},\
  }\href {\doibase 10.1103/PhysRevD.84.024020} {\bibfield  {journal} {\bibinfo
  {journal} {\prd}\ }\textbf {\bibinfo {volume} {84}},\ \bibinfo {eid} {024020}
  (\bibinfo {year} {2011})},\ \Eprint {http://arxiv.org/abs/1104.2669}
  {arXiv:1104.2669 [gr-qc]} \BibitemShut {NoStop}%
\bibitem [{\citenamefont {{Harko}}, \citenamefont {{Lobo}},\ and\ \citenamefont
  {{Minazzoli}}(2013)}]{Harko3}%
  \BibitemOpen
  \bibfield  {author} {\bibinfo {author} {\bibfnamefont {T.}~\bibnamefont
  {{Harko}}}, \bibinfo {author} {\bibfnamefont {F.~S.~N.}\ \bibnamefont
  {{Lobo}}}, \ and\ \bibinfo {author} {\bibfnamefont {O.}~\bibnamefont
  {{Minazzoli}}},\ }\href {\doibase 10.1103/PhysRevD.87.047501} {\bibfield
  {journal} {\bibinfo  {journal} {\prd}\ }\textbf {\bibinfo {volume} {87}},\
  \bibinfo {eid} {047501} (\bibinfo {year} {2013})},\ \Eprint
  {http://arxiv.org/abs/1210.4218} {arXiv:1210.4218 [gr-qc]} \BibitemShut
  {NoStop}%
\bibitem [{\citenamefont {{Harko}}\ \emph {et~al.}(2014)\citenamefont
  {{Harko}}, \citenamefont {{Lobo}}, \citenamefont {{Otalora}},\ and\
  \citenamefont {{Saridakis}}}]{Harko4}%
  \BibitemOpen
  \bibfield  {author} {\bibinfo {author} {\bibfnamefont {T.}~\bibnamefont
  {{Harko}}}, \bibinfo {author} {\bibfnamefont {F.~S.~N.}\ \bibnamefont
  {{Lobo}}}, \bibinfo {author} {\bibfnamefont {G.}~\bibnamefont {{Otalora}}}, \
  and\ \bibinfo {author} {\bibfnamefont {E.~N.}\ \bibnamefont {{Saridakis}}},\
  }\href {\doibase 10.1103/PhysRevD.89.124036} {\bibfield  {journal} {\bibinfo
  {journal} {\prd}\ }\textbf {\bibinfo {volume} {89}},\ \bibinfo {eid} {124036}
  (\bibinfo {year} {2014})},\ \Eprint {http://arxiv.org/abs/1404.6212}
  {arXiv:1404.6212 [gr-qc]} \BibitemShut {NoStop}%
\bibitem [{\citenamefont {{Lobo}}\ and\ \citenamefont
  {{Harko}}(2012)}]{harko-lobo-book}%
  \BibitemOpen
  \bibfield  {author} {\bibinfo {author} {\bibfnamefont {F.~S.~N.}\
  \bibnamefont {{Lobo}}}\ and\ \bibinfo {author} {\bibfnamefont
  {T.}~\bibnamefont {{Harko}}},\ }\href@noop {} {\bibfield  {journal} {\bibinfo
   {journal} {ArXiv e-prints}\ } (\bibinfo {year} {2012})},\ \Eprint
  {http://arxiv.org/abs/1211.0426} {arXiv:1211.0426 [gr-qc]} \BibitemShut
  {NoStop}%
\bibitem [{\citenamefont {{Bertolami}}\ \emph {et~al.}(2007)\citenamefont
  {{Bertolami}}, \citenamefont {{B{\"o}hmer}}, \citenamefont {{Harko}},\ and\
  \citenamefont {{Lobo}}}]{Bertolami}%
  \BibitemOpen
  \bibfield  {author} {\bibinfo {author} {\bibfnamefont {O.}~\bibnamefont
  {{Bertolami}}}, \bibinfo {author} {\bibfnamefont {C.~G.}\ \bibnamefont
  {{B{\"o}hmer}}}, \bibinfo {author} {\bibfnamefont {T.}~\bibnamefont
  {{Harko}}}, \ and\ \bibinfo {author} {\bibfnamefont {F.~S.~N.}\ \bibnamefont
  {{Lobo}}},\ }\href {\doibase 10.1103/PhysRevD.75.104016} {\bibfield
  {journal} {\bibinfo  {journal} {\prd}\ }\textbf {\bibinfo {volume} {75}},\
  \bibinfo {eid} {104016} (\bibinfo {year} {2007})},\ \Eprint
  {http://arxiv.org/abs/0704.1733} {arXiv:0704.1733 [gr-qc]} \BibitemShut
  {NoStop}%
\bibitem [{\citenamefont {{Barrientos}}\ and\ \citenamefont
  {{Mendoza}}(2018)}]{barrientos18}%
  \BibitemOpen
  \bibfield  {author} {\bibinfo {author} {\bibfnamefont {E.}~\bibnamefont
  {{Barrientos}}}\ and\ \bibinfo {author} {\bibfnamefont {S.}~\bibnamefont
  {{Mendoza}}},\ }\href {\doibase 10.1103/PhysRevD.98.084033} {\bibfield
  {journal} {\bibinfo  {journal} {\prd}\ }\textbf {\bibinfo {volume} {98}},\
  \bibinfo {eid} {084033} (\bibinfo {year} {2018})},\ \Eprint
  {http://arxiv.org/abs/1808.01386} {arXiv:1808.01386 [gr-qc]} \BibitemShut
  {NoStop}%
\bibitem [{\citenamefont {{Bernal}}\ \emph
  {et~al.}(2011{\natexlab{a}})\citenamefont {{Bernal}}, \citenamefont
  {{Capozziello}}, \citenamefont {{Hidalgo}},\ and\ \citenamefont
  {{Mendoza}}}]{bernal11}%
  \BibitemOpen
  \bibfield  {author} {\bibinfo {author} {\bibfnamefont {T.}~\bibnamefont
  {{Bernal}}}, \bibinfo {author} {\bibfnamefont {S.}~\bibnamefont
  {{Capozziello}}}, \bibinfo {author} {\bibfnamefont {J.~C.}\ \bibnamefont
  {{Hidalgo}}}, \ and\ \bibinfo {author} {\bibfnamefont {S.}~\bibnamefont
  {{Mendoza}}},\ }\href {\doibase 10.1140/epjc/s10052-011-1794-z} {\bibfield
  {journal} {\bibinfo  {journal} {European Physical Journal C}\ }\textbf
  {\bibinfo {volume} {71}},\ \bibinfo {eid} {1794} (\bibinfo {year}
  {2011}{\natexlab{a}})},\ \Eprint {http://arxiv.org/abs/1108.5588}
  {arXiv:1108.5588 [astro-ph.CO]} \BibitemShut {NoStop}%
\bibitem [{\citenamefont {{Mendoza}}\ \emph {et~al.}(2013)\citenamefont
  {{Mendoza}}, \citenamefont {{Bernal}}, \citenamefont {{Hernandez}},
  \citenamefont {{Hidalgo}},\ and\ \citenamefont {{Torres}}}]{mendoza13}%
  \BibitemOpen
  \bibfield  {author} {\bibinfo {author} {\bibfnamefont {S.}~\bibnamefont
  {{Mendoza}}}, \bibinfo {author} {\bibfnamefont {T.}~\bibnamefont {{Bernal}}},
  \bibinfo {author} {\bibfnamefont {X.}~\bibnamefont {{Hernandez}}}, \bibinfo
  {author} {\bibfnamefont {J.~C.}\ \bibnamefont {{Hidalgo}}}, \ and\ \bibinfo
  {author} {\bibfnamefont {L.~A.}\ \bibnamefont {{Torres}}},\ }\href {\doibase
  10.1093/mnras/stt752} {\bibfield  {journal} {\bibinfo  {journal} {\mnras}\
  }\textbf {\bibinfo {volume} {433}},\ \bibinfo {pages} {1802} (\bibinfo {year}
  {2013})},\ \Eprint {http://arxiv.org/abs/1208.6241} {arXiv:1208.6241
  [astro-ph.CO]} \BibitemShut {NoStop}%
\bibitem [{\citenamefont {{Barrientos}}\ and\ \citenamefont
  {{Mendoza}}(2016)}]{barrientos16}%
  \BibitemOpen
  \bibfield  {author} {\bibinfo {author} {\bibfnamefont {E.}~\bibnamefont
  {{Barrientos}}}\ and\ \bibinfo {author} {\bibfnamefont {S.}~\bibnamefont
  {{Mendoza}}},\ }\href {\doibase 10.1140/epjp/i2016-16367-0} {\bibfield
  {journal} {\bibinfo  {journal} {European Physical Journal Plus}\ }\textbf
  {\bibinfo {volume} {131}},\ \bibinfo {eid} {367} (\bibinfo {year} {2016})},\
  \Eprint {http://arxiv.org/abs/1602.05644} {arXiv:1602.05644 [gr-qc]}
  \BibitemShut {NoStop}%
\bibitem [{\citenamefont {{Barrientos}}, \citenamefont {{Bernal}},\ and\
  \citenamefont {{Mendoza}}(2020)}]{Barrientos20}%
  \BibitemOpen
  \bibfield  {author} {\bibinfo {author} {\bibfnamefont {E.}~\bibnamefont
  {{Barrientos}}}, \bibinfo {author} {\bibfnamefont {T.}~\bibnamefont
  {{Bernal}}}, \ and\ \bibinfo {author} {\bibfnamefont {S.}~\bibnamefont
  {{Mendoza}}},\ }\href@noop {} {\bibfield  {journal} {\bibinfo  {journal}
  {arXiv e-prints}\ ,\ \bibinfo {eid} {arXiv:2008.01800}} (\bibinfo {year}
  {2020})},\ \Eprint {http://arxiv.org/abs/2008.01800} {arXiv:2008.01800
  [gr-qc]} \BibitemShut {NoStop}%
\bibitem [{\citenamefont {{Mashhoon}}(1993)}]{Mash1}%
  \BibitemOpen
  \bibfield  {author} {\bibinfo {author} {\bibfnamefont {B.}~\bibnamefont
  {{Mashhoon}}},\ }\href {\doibase 10.1103/PhysRevA.47.4498} {\bibfield
  {journal} {\bibinfo  {journal} {\pra}\ }\textbf {\bibinfo {volume} {47}},\
  \bibinfo {pages} {4498} (\bibinfo {year} {1993})}\BibitemShut {NoStop}%
\bibitem [{\citenamefont {{Mashhoon}}(2001)}]{Mash2}%
  \BibitemOpen
  \bibfield  {author} {\bibinfo {author} {\bibfnamefont {B.}~\bibnamefont
  {{Mashhoon}}},\ }\href@noop {} {\bibfield  {journal} {\bibinfo  {journal}
  {arXiv e-prints}\ ,\ \bibinfo {eid} {gr-qc/0112058}} (\bibinfo {year}
  {2001})},\ \Eprint {http://arxiv.org/abs/gr-qc/0112058} {arXiv:gr-qc/0112058
  [gr-qc]} \BibitemShut {NoStop}%
\bibitem [{\citenamefont {{Chicone}}\ and\ \citenamefont
  {{Mashhoon}}(2012)}]{Chicone1}%
  \BibitemOpen
  \bibfield  {author} {\bibinfo {author} {\bibfnamefont {C.}~\bibnamefont
  {{Chicone}}}\ and\ \bibinfo {author} {\bibfnamefont {B.}~\bibnamefont
  {{Mashhoon}}},\ }\href {\doibase 10.1063/1.3702449} {\bibfield  {journal}
  {\bibinfo  {journal} {Journal of Mathematical Physics}\ }\textbf {\bibinfo
  {volume} {53}},\ \bibinfo {pages} {042501} (\bibinfo {year} {2012})},\
  \Eprint {http://arxiv.org/abs/1111.4702} {arXiv:1111.4702 [gr-qc]}
  \BibitemShut {NoStop}%
\bibitem [{\citenamefont {{Chicone}}\ and\ \citenamefont
  {{Mashhoon}}(2016{\natexlab{a}})}]{Chicone2}%
  \BibitemOpen
  \bibfield  {author} {\bibinfo {author} {\bibfnamefont {C.}~\bibnamefont
  {{Chicone}}}\ and\ \bibinfo {author} {\bibfnamefont {B.}~\bibnamefont
  {{Mashhoon}}},\ }\href {\doibase 10.1088/0264-9381/33/7/075005} {\bibfield
  {journal} {\bibinfo  {journal} {Classical and Quantum Gravity}\ }\textbf
  {\bibinfo {volume} {33}},\ \bibinfo {eid} {075005} (\bibinfo {year}
  {2016}{\natexlab{a}})},\ \Eprint {http://arxiv.org/abs/1508.01508}
  {arXiv:1508.01508 [gr-qc]} \BibitemShut {NoStop}%
\bibitem [{\citenamefont {{Chicone}}\ and\ \citenamefont
  {{Mashhoon}}(2016{\natexlab{b}})}]{Chicone3}%
  \BibitemOpen
  \bibfield  {author} {\bibinfo {author} {\bibfnamefont {C.}~\bibnamefont
  {{Chicone}}}\ and\ \bibinfo {author} {\bibfnamefont {B.}~\bibnamefont
  {{Mashhoon}}},\ }\href {\doibase 10.1063/1.4958902} {\bibfield  {journal}
  {\bibinfo  {journal} {Journal of Mathematical Physics}\ }\textbf {\bibinfo
  {volume} {57}},\ \bibinfo {eid} {072501} (\bibinfo {year}
  {2016}{\natexlab{b}})},\ \Eprint {http://arxiv.org/abs/1510.07316}
  {arXiv:1510.07316 [gr-qc]} \BibitemShut {NoStop}%
\bibitem [{\citenamefont {{Blome}}\ \emph {et~al.}(2010)\citenamefont
  {{Blome}}, \citenamefont {{Chicone}}, \citenamefont {{Hehl}},\ and\
  \citenamefont {{Mashhoon}}}]{Blome}%
  \BibitemOpen
  \bibfield  {author} {\bibinfo {author} {\bibfnamefont {H.-J.}\ \bibnamefont
  {{Blome}}}, \bibinfo {author} {\bibfnamefont {C.}~\bibnamefont {{Chicone}}},
  \bibinfo {author} {\bibfnamefont {F.~W.}\ \bibnamefont {{Hehl}}}, \ and\
  \bibinfo {author} {\bibfnamefont {B.}~\bibnamefont {{Mashhoon}}},\ }\href
  {\doibase 10.1103/PhysRevD.81.065020} {\bibfield  {journal} {\bibinfo
  {journal} {\prd}\ }\textbf {\bibinfo {volume} {81}},\ \bibinfo {eid} {065020}
  (\bibinfo {year} {2010})},\ \Eprint {http://arxiv.org/abs/1002.1425}
  {arXiv:1002.1425 [gr-qc]} \BibitemShut {NoStop}%
\bibitem [{\citenamefont {{Maggiore}}\ and\ \citenamefont
  {{Mancarella}}(2014)}]{Maggiore1}%
  \BibitemOpen
  \bibfield  {author} {\bibinfo {author} {\bibfnamefont {M.}~\bibnamefont
  {{Maggiore}}}\ and\ \bibinfo {author} {\bibfnamefont {M.}~\bibnamefont
  {{Mancarella}}},\ }\href {\doibase 10.1103/PhysRevD.90.023005} {\bibfield
  {journal} {\bibinfo  {journal} {\prd}\ }\textbf {\bibinfo {volume} {90}},\
  \bibinfo {eid} {023005} (\bibinfo {year} {2014})},\ \Eprint
  {http://arxiv.org/abs/1402.0448} {arXiv:1402.0448 [hep-th]} \BibitemShut
  {NoStop}%
\bibitem [{\citenamefont {{Hehl}}\ and\ \citenamefont
  {{Mashhoon}}(2009)}]{HehlM}%
  \BibitemOpen
  \bibfield  {author} {\bibinfo {author} {\bibfnamefont {F.~W.}\ \bibnamefont
  {{Hehl}}}\ and\ \bibinfo {author} {\bibfnamefont {B.}~\bibnamefont
  {{Mashhoon}}},\ }\href {\doibase 10.1016/j.physletb.2009.02.033} {\bibfield
  {journal} {\bibinfo  {journal} {Physics Letters B}\ }\textbf {\bibinfo
  {volume} {673}},\ \bibinfo {pages} {279} (\bibinfo {year} {2009})},\ \Eprint
  {http://arxiv.org/abs/0812.1059} {arXiv:0812.1059 [gr-qc]} \BibitemShut
  {NoStop}%
\bibitem [{\citenamefont {{Foffa}}, \citenamefont {{Maggiore}},\ and\
  \citenamefont {{Mitsou}}(2014)}]{Foffa}%
  \BibitemOpen
  \bibfield  {author} {\bibinfo {author} {\bibfnamefont {S.}~\bibnamefont
  {{Foffa}}}, \bibinfo {author} {\bibfnamefont {M.}~\bibnamefont {{Maggiore}}},
  \ and\ \bibinfo {author} {\bibfnamefont {E.}~\bibnamefont {{Mitsou}}},\
  }\href {\doibase 10.1142/S0217751X14501164} {\bibfield  {journal} {\bibinfo
  {journal} {International Journal of Modern Physics A}\ }\textbf {\bibinfo
  {volume} {29}},\ \bibinfo {eid} {1450116} (\bibinfo {year} {2014})},\ \Eprint
  {http://arxiv.org/abs/1311.3435} {arXiv:1311.3435 [hep-th]} \BibitemShut
  {NoStop}%
\bibitem [{\citenamefont {Podlubny}(1999)}]{Podlubny}%
  \BibitemOpen
  \bibfield  {author} {\bibinfo {author} {\bibfnamefont {I.}~\bibnamefont
  {Podlubny}},\ }\href {https://cds.cern.ch/record/395913} {\emph {\bibinfo
  {title} {{Fractional differential equations: an introduction to fractional
  derivatives, fractional differential equations, to methods of their solution
  and some of their applications}}}},\ Mathematics in Science and Engineering\
  (\bibinfo  {publisher} {Academic Press},\ \bibinfo {address} {London},\
  \bibinfo {year} {1999})\BibitemShut {NoStop}%
\bibitem [{\citenamefont {Kochubei}\ and\ \citenamefont
  {Luchko}(2019{\natexlab{a}})}]{Handbook1}%
  \BibitemOpen
  \bibfield  {author} {\bibinfo {author} {\bibfnamefont {A.}~\bibnamefont
  {Kochubei}}\ and\ \bibinfo {author} {\bibfnamefont {Y.}~\bibnamefont
  {Luchko}},\ }\href {\doibase https://doi.org/10.1515/9783110571622} {\emph
  {\bibinfo {title} {Basic Theory}}}\ (\bibinfo  {publisher} {De Gruyter},\
  \bibinfo {address} {Berlin, Boston},\ \bibinfo {year} {19 Feb.
  2019})\BibitemShut {NoStop}%
\bibitem [{\citenamefont {Kochubei}\ and\ \citenamefont
  {Luchko}(2019{\natexlab{b}})}]{Handbook2}%
  \BibitemOpen
  \bibfield  {author} {\bibinfo {author} {\bibfnamefont {A.}~\bibnamefont
  {Kochubei}}\ and\ \bibinfo {author} {\bibfnamefont {Y.}~\bibnamefont
  {Luchko}},\ }\href {\doibase https://doi.org/10.1515/9783110571660} {\emph
  {\bibinfo {title} {Fractional Differential Equations}}}\ (\bibinfo
  {publisher} {De Gruyter},\ \bibinfo {address} {Berlin, Boston},\ \bibinfo
  {year} {19 Feb. 2019})\BibitemShut {NoStop}%
\bibitem [{\citenamefont {Karniadakis}(2019)}]{Handbook3}%
  \BibitemOpen
  \bibfield  {author} {\bibinfo {author} {\bibfnamefont {G.~E.}\ \bibnamefont
  {Karniadakis}},\ }\href {\doibase https://doi.org/10.1515/9783110571684}
  {\emph {\bibinfo {title} {Numerical Methods}}}\ (\bibinfo  {publisher} {De
  Gruyter},\ \bibinfo {address} {Berlin, Boston},\ \bibinfo {year} {15 Apr.
  2019})\BibitemShut {NoStop}%
\bibitem [{\citenamefont {Tarasov}(2019)}]{Handbook4}%
  \BibitemOpen
  \bibfield  {author} {\bibinfo {author} {\bibfnamefont {V.~E.}\ \bibnamefont
  {Tarasov}},\ }\href {\doibase https://doi.org/10.1515/9783110571707} {\emph
  {\bibinfo {title} {Applications in Physics, Part A}}}\ (\bibinfo  {publisher}
  {De Gruyter},\ \bibinfo {address} {Berlin, Boston},\ \bibinfo {year} {19 Feb.
  2019})\BibitemShut {NoStop}%
\bibitem [{\citenamefont {Petráš}(2019)}]{Handbook6}%
  \BibitemOpen
  \bibfield  {author} {\bibinfo {author} {\bibfnamefont {I.}~\bibnamefont
  {Petráš}},\ }\href {\doibase https://doi.org/10.1515/9783110571745} {\emph
  {\bibinfo {title} {Applications in Control}}}\ (\bibinfo  {publisher} {De
  Gruyter},\ \bibinfo {address} {Berlin, Boston},\ \bibinfo {year} {19 Feb.
  2019})\BibitemShut {NoStop}%
\bibitem [{\citenamefont {Bǎleanu}\ and\ \citenamefont
  {Lopes}(2019)}]{Handbook7}%
  \BibitemOpen
  \bibfield  {author} {\bibinfo {author} {\bibfnamefont {D.}~\bibnamefont
  {Bǎleanu}}\ and\ \bibinfo {author} {\bibfnamefont {A.~M.}\ \bibnamefont
  {Lopes}},\ }\href {\doibase https://doi.org/10.1515/9783110571905} {\emph
  {\bibinfo {title} {Applications in Engineering, Life and Social Sciences,
  Part A}}}\ (\bibinfo  {publisher} {De Gruyter},\ \bibinfo {address} {Berlin,
  Boston},\ \bibinfo {year} {01 Apr. 2019})\BibitemShut {NoStop}%
\bibitem [{\citenamefont {{Shchigolev}}(2011)}]{shch1}%
  \BibitemOpen
  \bibfield  {author} {\bibinfo {author} {\bibfnamefont {V.~K.}\ \bibnamefont
  {{Shchigolev}}},\ }\href {\doibase 10.1088/0253-6102/56/2/34} {\bibfield
  {journal} {\bibinfo  {journal} {Communications in Theoretical Physics}\
  }\textbf {\bibinfo {volume} {56}},\ \bibinfo {pages} {389} (\bibinfo {year}
  {2011})},\ \Eprint {http://arxiv.org/abs/1011.3304} {arXiv:1011.3304 [gr-qc]}
  \BibitemShut {NoStop}%
\bibitem [{\citenamefont {{Shchigolev}}(2013{\natexlab{a}})}]{shch2}%
  \BibitemOpen
  \bibfield  {author} {\bibinfo {author} {\bibfnamefont {V.~K.}\ \bibnamefont
  {{Shchigolev}}},\ }\href {\doibase 10.5890/DNC.2013.04.002} {\bibfield
  {journal} {\bibinfo  {journal} {Discontinuitynlinearity and Complexity}\
  }\textbf {\bibinfo {volume} {2}},\ \bibinfo {pages} {115} (\bibinfo {year}
  {2013}{\natexlab{a}})},\ \Eprint {http://arxiv.org/abs/1208.3454}
  {arXiv:1208.3454 [gr-qc]} \BibitemShut {NoStop}%
\bibitem [{\citenamefont {{Shchigolev}}(2016)}]{shch3}%
  \BibitemOpen
  \bibfield  {author} {\bibinfo {author} {\bibfnamefont {V.~K.}\ \bibnamefont
  {{Shchigolev}}},\ }\href {\doibase 10.1140/epjp/i2016-16256-6} {\bibfield
  {journal} {\bibinfo  {journal} {European Physical Journal Plus}\ }\textbf
  {\bibinfo {volume} {131}},\ \bibinfo {eid} {256} (\bibinfo {year} {2016})},\
  \Eprint {http://arxiv.org/abs/1512.04113} {arXiv:1512.04113 [gr-qc]}
  \BibitemShut {NoStop}%
\bibitem [{\citenamefont {{Shchigolev}}(2013{\natexlab{b}})}]{shch4}%
  \BibitemOpen
  \bibfield  {author} {\bibinfo {author} {\bibfnamefont {V.~K.}\ \bibnamefont
  {{Shchigolev}}},\ }\href {\doibase 10.1142/S0217732313500569} {\bibfield
  {journal} {\bibinfo  {journal} {Modern Physics Letters A}\ }\textbf {\bibinfo
  {volume} {28}},\ \bibinfo {eid} {1350056} (\bibinfo {year}
  {2013}{\natexlab{b}})},\ \Eprint {http://arxiv.org/abs/1301.7198}
  {arXiv:1301.7198 [gr-qc]} \BibitemShut {NoStop}%
\bibitem [{\citenamefont {{Roberts}}(2009)}]{Roberts}%
  \BibitemOpen
  \bibfield  {author} {\bibinfo {author} {\bibfnamefont {M.~D.}\ \bibnamefont
  {{Roberts}}},\ }\href@noop {} {\bibfield  {journal} {\bibinfo  {journal}
  {arXiv e-prints}\ ,\ \bibinfo {eid} {arXiv:0909.1171}} (\bibinfo {year}
  {2009})},\ \Eprint {http://arxiv.org/abs/0909.1171} {arXiv:0909.1171 [gr-qc]}
  \BibitemShut {NoStop}%
\bibitem [{\citenamefont {{Vacaru}}(2012)}]{Vacaru}%
  \BibitemOpen
  \bibfield  {author} {\bibinfo {author} {\bibfnamefont {S.~I.}\ \bibnamefont
  {{Vacaru}}},\ }\href {\doibase 10.1007/s10773-011-1010-9} {\bibfield
  {journal} {\bibinfo  {journal} {International Journal of Theoretical
  Physics}\ }\textbf {\bibinfo {volume} {51}},\ \bibinfo {pages} {1338}
  (\bibinfo {year} {2012})},\ \Eprint {http://arxiv.org/abs/1004.0628}
  {arXiv:1004.0628 [math-ph]} \BibitemShut {NoStop}%
\bibitem [{\citenamefont {{Rami}}(2005)}]{Rami}%
  \BibitemOpen
  \bibfield  {author} {\bibinfo {author} {\bibfnamefont {E.-N.~A.}\
  \bibnamefont {{Rami}}},\ }\href@noop {} {\bibfield  {journal} {\bibinfo
  {journal} {Electronic Journal of Theoretical Physics}\ }\textbf {\bibinfo
  {volume} {2}},\ \bibinfo {pages} {1} (\bibinfo {year} {2005})}\BibitemShut
  {NoStop}%
\bibitem [{\citenamefont {{Frederico}}\ and\ \citenamefont
  {{Torres}}(2006)}]{Frederico}%
  \BibitemOpen
  \bibfield  {author} {\bibinfo {author} {\bibfnamefont {G.~S.~F.}\
  \bibnamefont {{Frederico}}}\ and\ \bibinfo {author} {\bibfnamefont
  {D.~F.~M.}\ \bibnamefont {{Torres}}},\ }\href@noop {} {\bibfield  {journal}
  {\bibinfo  {journal} {arXiv Mathematics e-prints}\ ,\ \bibinfo {eid}
  {math/0607472}} (\bibinfo {year} {2006})},\ \Eprint
  {http://arxiv.org/abs/math/0607472} {arXiv:math/0607472 [math.OC]}
  \BibitemShut {NoStop}%
\bibitem [{\citenamefont {Baleanu}(2007)}]{Balenu07}%
  \BibitemOpen
  \bibfield  {author} {\bibinfo {author} {\bibfnamefont {D.}~\bibnamefont
  {Baleanu}},\ }\href {\doibase 10.1002/pamm.200700327} {\bibfield  {journal}
  {\bibinfo  {journal} {Proceedings in Applied Mathematics and Mechanics}\
  }\textbf {\bibinfo {volume} {7}} (\bibinfo {year} {2007}),\
  10.1002/pamm.200700327}\BibitemShut {NoStop}%
\bibitem [{\citenamefont {{El-Nabulsi}}\ and\ \citenamefont
  {{Torres}}(2008)}]{Nabulsi}%
  \BibitemOpen
  \bibfield  {author} {\bibinfo {author} {\bibfnamefont {R.~A.}\ \bibnamefont
  {{El-Nabulsi}}}\ and\ \bibinfo {author} {\bibfnamefont {D.~F.~M.}\
  \bibnamefont {{Torres}}},\ }\href {\doibase 10.1063/1.2929662} {\bibfield
  {journal} {\bibinfo  {journal} {Journal of Mathematical Physics}\ }\textbf
  {\bibinfo {volume} {49}},\ \bibinfo {eid} {053521} (\bibinfo {year}
  {2008})},\ \Eprint {http://arxiv.org/abs/0804.4500} {arXiv:0804.4500
  [math-ph]} \BibitemShut {NoStop}%
\bibitem [{\citenamefont {Herzallah}\ and\ \citenamefont
  {Baleanu}(2009)}]{Herzallah}%
  \BibitemOpen
  \bibfield  {author} {\bibinfo {author} {\bibfnamefont {M.}~\bibnamefont
  {Herzallah}}\ and\ \bibinfo {author} {\bibfnamefont {D.}~\bibnamefont
  {Baleanu}},\ }\href {\doibase 10.1007/s11071-009-9486-z} {\bibfield
  {journal} {\bibinfo  {journal} {Nonlinear Dynamics}\ }\textbf {\bibinfo
  {volume} {58}},\ \bibinfo {pages} {385} (\bibinfo {year} {2009})}\BibitemShut
  {NoStop}%
\bibitem [{\citenamefont {Baleanu}\ and\ \citenamefont
  {Muslih}(2005)}]{Baleanu05}%
  \BibitemOpen
  \bibfield  {author} {\bibinfo {author} {\bibfnamefont {D.}~\bibnamefont
  {Baleanu}}\ and\ \bibinfo {author} {\bibfnamefont {S.~I.}\ \bibnamefont
  {Muslih}},\ }\href {\doibase 10.1238/physica.regular.072a00119} {\bibfield
  {journal} {\bibinfo  {journal} {Physica Scripta}\ }\textbf {\bibinfo {volume}
  {72}},\ \bibinfo {pages} {119} (\bibinfo {year} {2005})}\BibitemShut
  {NoStop}%
\bibitem [{\citenamefont {El-Nabulsi}(2013)}]{Nabusi13}%
  \BibitemOpen
  \bibfield  {author} {\bibinfo {author} {\bibfnamefont {R.}~\bibnamefont
  {El-Nabulsi}},\ }\href {\doibase 10.1007/s11071-013-0977-6} {\bibfield
  {journal} {\bibinfo  {journal} {Nonlinear Dynamics}\ }\textbf {\bibinfo
  {volume} {74}} (\bibinfo {year} {2013}),\
  10.1007/s11071-013-0977-6}\BibitemShut {NoStop}%
\bibitem [{\citenamefont {{Peacock}}(1999)}]{Peacock}%
  \BibitemOpen
  \bibfield  {author} {\bibinfo {author} {\bibfnamefont {J.~A.}\ \bibnamefont
  {{Peacock}}},\ }\href@noop {} {\emph {\bibinfo {title} {{Cosmological
  Physics}}}}\ (\bibinfo {year} {1999})\BibitemShut {NoStop}%
\bibitem [{\citenamefont {{Misner}}, \citenamefont {{Thorne}},\ and\
  \citenamefont {{Wheeler}}(1973)}]{Misner}%
  \BibitemOpen
  \bibfield  {author} {\bibinfo {author} {\bibfnamefont {C.~W.}\ \bibnamefont
  {{Misner}}}, \bibinfo {author} {\bibfnamefont {K.~S.}\ \bibnamefont
  {{Thorne}}}, \ and\ \bibinfo {author} {\bibfnamefont {J.~A.}\ \bibnamefont
  {{Wheeler}}},\ }\href@noop {} {\emph {\bibinfo {title} {{Gravitation}}}}\
  (\bibinfo {year} {1973})\BibitemShut {NoStop}%
\bibitem [{\citenamefont {{Longair}}(1989)}]{Longair}%
  \BibitemOpen
  \bibfield  {author} {\bibinfo {author} {\bibfnamefont {M.~S.}\ \bibnamefont
  {{Longair}}},\ }\enquote {\bibinfo {title} {{Galaxy Formation}},}\ in\ \href
  {\doibase 10.1007/3-540-51315-9_1} {\emph {\bibinfo {booktitle} {Evolution of
  Galaxies: Astronomical Observations}}},\ Vol.\ \bibinfo {volume} {333},\
  \bibinfo {editor} {edited by\ \bibinfo {editor} {\bibfnamefont
  {I.}~\bibnamefont {{Appenzeller}}}, \bibinfo {editor} {\bibfnamefont {H.~J.}\
  \bibnamefont {{Habing}}}, \ and\ \bibinfo {editor} {\bibfnamefont
  {P.}~\bibnamefont {{Lena}}}}\ (\bibinfo {year} {1989})\ p.~\bibinfo {pages}
  {1}\BibitemShut {NoStop}%
\bibitem [{\citenamefont {{Peebles}}(1993)}]{Peebles}%
  \BibitemOpen
  \bibfield  {author} {\bibinfo {author} {\bibfnamefont {P.~J.~E.}\
  \bibnamefont {{Peebles}}},\ }\href@noop {} {\emph {\bibinfo {title}
  {{Principles of Physical Cosmology}}}}\ (\bibinfo {year} {1993})\BibitemShut
  {NoStop}%
\bibitem [{\citenamefont {{Li}}, \citenamefont {{Du}},\ and\ \citenamefont
  {{Xu}}(2020)}]{Kun}%
  \BibitemOpen
  \bibfield  {author} {\bibinfo {author} {\bibfnamefont {E.-K.}\ \bibnamefont
  {{Li}}}, \bibinfo {author} {\bibfnamefont {M.}~\bibnamefont {{Du}}}, \ and\
  \bibinfo {author} {\bibfnamefont {L.}~\bibnamefont {{Xu}}},\ }\href {\doibase
  10.1093/mnras/stz3308} {\bibfield  {journal} {\bibinfo  {journal} {\mnras}\
  }\textbf {\bibinfo {volume} {491}},\ \bibinfo {pages} {4960} (\bibinfo {year}
  {2020})},\ \Eprint {http://arxiv.org/abs/1903.11433} {arXiv:1903.11433
  [astro-ph.CO]} \BibitemShut {NoStop}%
\bibitem [{\citenamefont {{Planck Collaboration}}\ \emph
  {et~al.}(2018)\citenamefont {{Planck Collaboration}}, \citenamefont
  {{Aghanim}}, \citenamefont {{Akrami}}, \citenamefont {{Ashdown}},
  \citenamefont {{Aumont}}, \citenamefont {{Baccigalupi}}, \citenamefont
  {{Ballardini}}, \citenamefont {{Banday}}, \citenamefont {{Barreiro}},
  \citenamefont {{Bartolo}}, \citenamefont {{Basak}}, \citenamefont {{Battye}},
  \citenamefont {{Benabed}}, \citenamefont {{Bernard}}, \citenamefont
  {{Bersanelli}}, \citenamefont {{Bielewicz}}, \citenamefont {{Bock}},
  \citenamefont {{Bond}}, \citenamefont {{Borrill}}, \citenamefont {{Bouchet}},
  \citenamefont {{Boulanger}}, \citenamefont {{Bucher}}, \citenamefont
  {{Burigana}}, \citenamefont {{Butler}}, \citenamefont {{Calabrese}},
  \citenamefont {{Cardoso}}, \citenamefont {{Carron}}, \citenamefont
  {{Challinor}}, \citenamefont {{Chiang}}, \citenamefont {{Chluba}},
  \citenamefont {{Colombo}}, \citenamefont {{Combet}}, \citenamefont
  {{Contreras}}, \citenamefont {{Crill}}, \citenamefont {{Cuttaia}},
  \citenamefont {{de Bernardis}}, \citenamefont {{de Zotti}}, \citenamefont
  {{Delabrouille}}, \citenamefont {{Delouis}}, \citenamefont {{Di Valentino}},
  \citenamefont {{Diego}}, \citenamefont {{Dor{\'e}}}, \citenamefont
  {{Douspis}}, \citenamefont {{Ducout}}, \citenamefont {{Dupac}}, \citenamefont
  {{Dusini}}, \citenamefont {{Efstathiou}}, \citenamefont {{Elsner}},
  \citenamefont {{En{\ss}lin}}, \citenamefont {{Eriksen}}, \citenamefont
  {{Fantaye}}, \citenamefont {{Farhang}}, \citenamefont {{Fergusson}},
  \citenamefont {{Fernandez-Cobos}}, \citenamefont {{Finelli}}, \citenamefont
  {{Forastieri}}, \citenamefont {{Frailis}}, \citenamefont {{Fraisse}},
  \citenamefont {{Franceschi}}, \citenamefont {{Frolov}}, \citenamefont
  {{Galeotta}}, \citenamefont {{Galli}}, \citenamefont {{Ganga}}, \citenamefont
  {{G{\'e}nova-Santos}}, \citenamefont {{Gerbino}}, \citenamefont {{Ghosh}},
  \citenamefont {{Gonz{\'a}lez-Nuevo}}, \citenamefont {{G{\'o}rski}},
  \citenamefont {{Gratton}}, \citenamefont {{Gruppuso}}, \citenamefont
  {{Gudmundsson}}, \citenamefont {{Hamann}}, \citenamefont {{Handley}},
  \citenamefont {{Hansen}}, \citenamefont {{Herranz}}, \citenamefont
  {{Hildebrandt}}, \citenamefont {{Hivon}}, \citenamefont {{Huang}},
  \citenamefont {{Jaffe}}, \citenamefont {{Jones}}, \citenamefont {{Karakci}},
  \citenamefont {{Keih{\"a}nen}}, \citenamefont {{Keskitalo}}, \citenamefont
  {{Kiiveri}}, \citenamefont {{Kim}}, \citenamefont {{Kisner}}, \citenamefont
  {{Knox}}, \citenamefont {{Krachmalnicoff}}, \citenamefont {{Kunz}},
  \citenamefont {{Kurki-Suonio}}, \citenamefont {{Lagache}}, \citenamefont
  {{Lamarre}}, \citenamefont {{Lasenby}}, \citenamefont {{Lattanzi}},
  \citenamefont {{Lawrence}}, \citenamefont {{Le Jeune}}, \citenamefont
  {{Lemos}}, \citenamefont {{Lesgourgues}}, \citenamefont {{Levrier}},
  \citenamefont {{Lewis}}, \citenamefont {{Liguori}}, \citenamefont {{Lilje}},
  \citenamefont {{Lilley}}, \citenamefont {{Lindholm}}, \citenamefont
  {{L{\'o}pez-Caniego}}, \citenamefont {{Lubin}}, \citenamefont {{Ma}},
  \citenamefont {{Mac{\'\i}as-P{\'e}rez}}, \citenamefont {{Maggio}},
  \citenamefont {{Maino}}, \citenamefont {{Mandolesi}}, \citenamefont
  {{Mangilli}}, \citenamefont {{Marcos-Caballero}}, \citenamefont {{Maris}},
  \citenamefont {{Martin}}, \citenamefont {{Martinelli}}, \citenamefont
  {{Mart{\'\i}nez-Gonz{\'a}lez}}, \citenamefont {{Matarrese}}, \citenamefont
  {{Mauri}}, \citenamefont {{McEwen}}, \citenamefont {{Meinhold}},
  \citenamefont {{Melchiorri}}, \citenamefont {{Mennella}}, \citenamefont
  {{Migliaccio}}, \citenamefont {{Millea}}, \citenamefont {{Mitra}},
  \citenamefont {{Miville-Desch{\^e}nes}}, \citenamefont {{Molinari}},
  \citenamefont {{Montier}}, \citenamefont {{Morgante}}, \citenamefont
  {{Moss}}, \citenamefont {{Natoli}}, \citenamefont {{N{\o}rgaard-Nielsen}},
  \citenamefont {{Pagano}}, \citenamefont {{Paoletti}}, \citenamefont
  {{Partridge}}, \citenamefont {{Patanchon}}, \citenamefont {{Peiris}},
  \citenamefont {{Perrotta}}, \citenamefont {{Pettorino}}, \citenamefont
  {{Piacentini}}, \citenamefont {{Polastri}}, \citenamefont {{Polenta}},
  \citenamefont {{Puget}}, \citenamefont {{Rachen}}, \citenamefont
  {{Reinecke}}, \citenamefont {{Remazeilles}}, \citenamefont {{Renzi}},
  \citenamefont {{Rocha}}, \citenamefont {{Rosset}}, \citenamefont {{Roudier}},
  \citenamefont {{Rubi{\~n}o-Mart{\'\i}n}}, \citenamefont {{Ruiz-Granados}},
  \citenamefont {{Salvati}}, \citenamefont {{Sandri}}, \citenamefont
  {{Savelainen}}, \citenamefont {{Scott}}, \citenamefont {{Shellard}},
  \citenamefont {{Sirignano}}, \citenamefont {{Sirri}}, \citenamefont
  {{Spencer}}, \citenamefont {{Sunyaev}}, \citenamefont {{Suur-Uski}},
  \citenamefont {{Tauber}}, \citenamefont {{Tavagnacco}}, \citenamefont
  {{Tenti}}, \citenamefont {{Toffolatti}}, \citenamefont {{Tomasi}},
  \citenamefont {{Trombetti}}, \citenamefont {{Valenziano}}, \citenamefont
  {{Valiviita}}, \citenamefont {{Van Tent}}, \citenamefont {{Vibert}},
  \citenamefont {{Vielva}}, \citenamefont {{Villa}}, \citenamefont
  {{Vittorio}}, \citenamefont {{Wand elt}}, \citenamefont {{Wehus}},
  \citenamefont {{White}}, \citenamefont {{White}}, \citenamefont {{Zacchei}},\
  and\ \citenamefont {{Zonca}}}]{planck}%
  \BibitemOpen
  \bibfield  {author} {\bibinfo {author} {\bibnamefont {{Planck
  Collaboration}}}, \bibinfo {author} {\bibfnamefont {N.}~\bibnamefont
  {{Aghanim}}}, \bibinfo {author} {\bibfnamefont {Y.}~\bibnamefont {{Akrami}}},
  \bibinfo {author} {\bibfnamefont {M.}~\bibnamefont {{Ashdown}}}, \bibinfo
  {author} {\bibfnamefont {J.}~\bibnamefont {{Aumont}}}, \bibinfo {author}
  {\bibfnamefont {C.}~\bibnamefont {{Baccigalupi}}}, \bibinfo {author}
  {\bibfnamefont {M.}~\bibnamefont {{Ballardini}}}, \bibinfo {author}
  {\bibfnamefont {A.~J.}\ \bibnamefont {{Banday}}}, \bibinfo {author}
  {\bibfnamefont {R.~B.}\ \bibnamefont {{Barreiro}}}, \bibinfo {author}
  {\bibfnamefont {N.}~\bibnamefont {{Bartolo}}}, \bibinfo {author}
  {\bibfnamefont {S.}~\bibnamefont {{Basak}}}, \bibinfo {author} {\bibfnamefont
  {R.}~\bibnamefont {{Battye}}}, \bibinfo {author} {\bibfnamefont
  {K.}~\bibnamefont {{Benabed}}}, \bibinfo {author} {\bibfnamefont {J.~P.}\
  \bibnamefont {{Bernard}}}, \bibinfo {author} {\bibfnamefont {M.}~\bibnamefont
  {{Bersanelli}}}, \bibinfo {author} {\bibfnamefont {P.}~\bibnamefont
  {{Bielewicz}}}, \bibinfo {author} {\bibfnamefont {J.~J.}\ \bibnamefont
  {{Bock}}}, \bibinfo {author} {\bibfnamefont {J.~R.}\ \bibnamefont {{Bond}}},
  \bibinfo {author} {\bibfnamefont {J.}~\bibnamefont {{Borrill}}}, \bibinfo
  {author} {\bibfnamefont {F.~R.}\ \bibnamefont {{Bouchet}}}, \bibinfo {author}
  {\bibfnamefont {F.}~\bibnamefont {{Boulanger}}}, \bibinfo {author}
  {\bibfnamefont {M.}~\bibnamefont {{Bucher}}}, \bibinfo {author}
  {\bibfnamefont {C.}~\bibnamefont {{Burigana}}}, \bibinfo {author}
  {\bibfnamefont {R.~C.}\ \bibnamefont {{Butler}}}, \bibinfo {author}
  {\bibfnamefont {E.}~\bibnamefont {{Calabrese}}}, \bibinfo {author}
  {\bibfnamefont {J.~F.}\ \bibnamefont {{Cardoso}}}, \bibinfo {author}
  {\bibfnamefont {J.}~\bibnamefont {{Carron}}}, \bibinfo {author}
  {\bibfnamefont {A.}~\bibnamefont {{Challinor}}}, \bibinfo {author}
  {\bibfnamefont {H.~C.}\ \bibnamefont {{Chiang}}}, \bibinfo {author}
  {\bibfnamefont {J.}~\bibnamefont {{Chluba}}}, \bibinfo {author}
  {\bibfnamefont {L.~P.~L.}\ \bibnamefont {{Colombo}}}, \bibinfo {author}
  {\bibfnamefont {C.}~\bibnamefont {{Combet}}}, \bibinfo {author}
  {\bibfnamefont {D.}~\bibnamefont {{Contreras}}}, \bibinfo {author}
  {\bibfnamefont {B.~P.}\ \bibnamefont {{Crill}}}, \bibinfo {author}
  {\bibfnamefont {F.}~\bibnamefont {{Cuttaia}}}, \bibinfo {author}
  {\bibfnamefont {P.}~\bibnamefont {{de Bernardis}}}, \bibinfo {author}
  {\bibfnamefont {G.}~\bibnamefont {{de Zotti}}}, \bibinfo {author}
  {\bibfnamefont {J.}~\bibnamefont {{Delabrouille}}}, \bibinfo {author}
  {\bibfnamefont {J.~M.}\ \bibnamefont {{Delouis}}}, \bibinfo {author}
  {\bibfnamefont {E.}~\bibnamefont {{Di Valentino}}}, \bibinfo {author}
  {\bibfnamefont {J.~M.}\ \bibnamefont {{Diego}}}, \bibinfo {author}
  {\bibfnamefont {O.}~\bibnamefont {{Dor{\'e}}}}, \bibinfo {author}
  {\bibfnamefont {M.}~\bibnamefont {{Douspis}}}, \bibinfo {author}
  {\bibfnamefont {A.}~\bibnamefont {{Ducout}}}, \bibinfo {author}
  {\bibfnamefont {X.}~\bibnamefont {{Dupac}}}, \bibinfo {author} {\bibfnamefont
  {S.}~\bibnamefont {{Dusini}}}, \bibinfo {author} {\bibfnamefont
  {G.}~\bibnamefont {{Efstathiou}}}, \bibinfo {author} {\bibfnamefont
  {F.}~\bibnamefont {{Elsner}}}, \bibinfo {author} {\bibfnamefont {T.~A.}\
  \bibnamefont {{En{\ss}lin}}}, \bibinfo {author} {\bibfnamefont {H.~K.}\
  \bibnamefont {{Eriksen}}}, \bibinfo {author} {\bibfnamefont {Y.}~\bibnamefont
  {{Fantaye}}}, \bibinfo {author} {\bibfnamefont {M.}~\bibnamefont
  {{Farhang}}}, \bibinfo {author} {\bibfnamefont {J.}~\bibnamefont
  {{Fergusson}}}, \bibinfo {author} {\bibfnamefont {R.}~\bibnamefont
  {{Fernandez-Cobos}}}, \bibinfo {author} {\bibfnamefont {F.}~\bibnamefont
  {{Finelli}}}, \bibinfo {author} {\bibfnamefont {F.}~\bibnamefont
  {{Forastieri}}}, \bibinfo {author} {\bibfnamefont {M.}~\bibnamefont
  {{Frailis}}}, \bibinfo {author} {\bibfnamefont {A.~A.}\ \bibnamefont
  {{Fraisse}}}, \bibinfo {author} {\bibfnamefont {E.}~\bibnamefont
  {{Franceschi}}}, \bibinfo {author} {\bibfnamefont {A.}~\bibnamefont
  {{Frolov}}}, \bibinfo {author} {\bibfnamefont {S.}~\bibnamefont
  {{Galeotta}}}, \bibinfo {author} {\bibfnamefont {S.}~\bibnamefont {{Galli}}},
  \bibinfo {author} {\bibfnamefont {K.}~\bibnamefont {{Ganga}}}, \bibinfo
  {author} {\bibfnamefont {R.~T.}\ \bibnamefont {{G{\'e}nova-Santos}}},
  \bibinfo {author} {\bibfnamefont {M.}~\bibnamefont {{Gerbino}}}, \bibinfo
  {author} {\bibfnamefont {T.}~\bibnamefont {{Ghosh}}}, \bibinfo {author}
  {\bibfnamefont {J.}~\bibnamefont {{Gonz{\'a}lez-Nuevo}}}, \bibinfo {author}
  {\bibfnamefont {K.~M.}\ \bibnamefont {{G{\'o}rski}}}, \bibinfo {author}
  {\bibfnamefont {S.}~\bibnamefont {{Gratton}}}, \bibinfo {author}
  {\bibfnamefont {A.}~\bibnamefont {{Gruppuso}}}, \bibinfo {author}
  {\bibfnamefont {J.~E.}\ \bibnamefont {{Gudmundsson}}}, \bibinfo {author}
  {\bibfnamefont {J.}~\bibnamefont {{Hamann}}}, \bibinfo {author}
  {\bibfnamefont {W.}~\bibnamefont {{Handley}}}, \bibinfo {author}
  {\bibfnamefont {F.~K.}\ \bibnamefont {{Hansen}}}, \bibinfo {author}
  {\bibfnamefont {D.}~\bibnamefont {{Herranz}}}, \bibinfo {author}
  {\bibfnamefont {S.~R.}\ \bibnamefont {{Hildebrandt}}}, \bibinfo {author}
  {\bibfnamefont {E.}~\bibnamefont {{Hivon}}}, \bibinfo {author} {\bibfnamefont
  {Z.}~\bibnamefont {{Huang}}}, \bibinfo {author} {\bibfnamefont {A.~H.}\
  \bibnamefont {{Jaffe}}}, \bibinfo {author} {\bibfnamefont {W.~C.}\
  \bibnamefont {{Jones}}}, \bibinfo {author} {\bibfnamefont {A.}~\bibnamefont
  {{Karakci}}}, \bibinfo {author} {\bibfnamefont {E.}~\bibnamefont
  {{Keih{\"a}nen}}}, \bibinfo {author} {\bibfnamefont {R.}~\bibnamefont
  {{Keskitalo}}}, \bibinfo {author} {\bibfnamefont {K.}~\bibnamefont
  {{Kiiveri}}}, \bibinfo {author} {\bibfnamefont {J.}~\bibnamefont {{Kim}}},
  \bibinfo {author} {\bibfnamefont {T.~S.}\ \bibnamefont {{Kisner}}}, \bibinfo
  {author} {\bibfnamefont {L.}~\bibnamefont {{Knox}}}, \bibinfo {author}
  {\bibfnamefont {N.}~\bibnamefont {{Krachmalnicoff}}}, \bibinfo {author}
  {\bibfnamefont {M.}~\bibnamefont {{Kunz}}}, \bibinfo {author} {\bibfnamefont
  {H.}~\bibnamefont {{Kurki-Suonio}}}, \bibinfo {author} {\bibfnamefont
  {G.}~\bibnamefont {{Lagache}}}, \bibinfo {author} {\bibfnamefont {J.~M.}\
  \bibnamefont {{Lamarre}}}, \bibinfo {author} {\bibfnamefont {A.}~\bibnamefont
  {{Lasenby}}}, \bibinfo {author} {\bibfnamefont {M.}~\bibnamefont
  {{Lattanzi}}}, \bibinfo {author} {\bibfnamefont {C.~R.}\ \bibnamefont
  {{Lawrence}}}, \bibinfo {author} {\bibfnamefont {M.}~\bibnamefont {{Le
  Jeune}}}, \bibinfo {author} {\bibfnamefont {P.}~\bibnamefont {{Lemos}}},
  \bibinfo {author} {\bibfnamefont {J.}~\bibnamefont {{Lesgourgues}}}, \bibinfo
  {author} {\bibfnamefont {F.}~\bibnamefont {{Levrier}}}, \bibinfo {author}
  {\bibfnamefont {A.}~\bibnamefont {{Lewis}}}, \bibinfo {author} {\bibfnamefont
  {M.}~\bibnamefont {{Liguori}}}, \bibinfo {author} {\bibfnamefont {P.~B.}\
  \bibnamefont {{Lilje}}}, \bibinfo {author} {\bibfnamefont {M.}~\bibnamefont
  {{Lilley}}}, \bibinfo {author} {\bibfnamefont {V.}~\bibnamefont
  {{Lindholm}}}, \bibinfo {author} {\bibfnamefont {M.}~\bibnamefont
  {{L{\'o}pez-Caniego}}}, \bibinfo {author} {\bibfnamefont {P.~M.}\
  \bibnamefont {{Lubin}}}, \bibinfo {author} {\bibfnamefont {Y.~Z.}\
  \bibnamefont {{Ma}}}, \bibinfo {author} {\bibfnamefont {J.~F.}\ \bibnamefont
  {{Mac{\'\i}as-P{\'e}rez}}}, \bibinfo {author} {\bibfnamefont
  {G.}~\bibnamefont {{Maggio}}}, \bibinfo {author} {\bibfnamefont
  {D.}~\bibnamefont {{Maino}}}, \bibinfo {author} {\bibfnamefont
  {N.}~\bibnamefont {{Mandolesi}}}, \bibinfo {author} {\bibfnamefont
  {A.}~\bibnamefont {{Mangilli}}}, \bibinfo {author} {\bibfnamefont
  {A.}~\bibnamefont {{Marcos-Caballero}}}, \bibinfo {author} {\bibfnamefont
  {M.}~\bibnamefont {{Maris}}}, \bibinfo {author} {\bibfnamefont {P.~G.}\
  \bibnamefont {{Martin}}}, \bibinfo {author} {\bibfnamefont {M.}~\bibnamefont
  {{Martinelli}}}, \bibinfo {author} {\bibfnamefont {E.}~\bibnamefont
  {{Mart{\'\i}nez-Gonz{\'a}lez}}}, \bibinfo {author} {\bibfnamefont
  {S.}~\bibnamefont {{Matarrese}}}, \bibinfo {author} {\bibfnamefont
  {N.}~\bibnamefont {{Mauri}}}, \bibinfo {author} {\bibfnamefont {J.~D.}\
  \bibnamefont {{McEwen}}}, \bibinfo {author} {\bibfnamefont {P.~R.}\
  \bibnamefont {{Meinhold}}}, \bibinfo {author} {\bibfnamefont
  {A.}~\bibnamefont {{Melchiorri}}}, \bibinfo {author} {\bibfnamefont
  {A.}~\bibnamefont {{Mennella}}}, \bibinfo {author} {\bibfnamefont
  {M.}~\bibnamefont {{Migliaccio}}}, \bibinfo {author} {\bibfnamefont
  {M.}~\bibnamefont {{Millea}}}, \bibinfo {author} {\bibfnamefont
  {S.}~\bibnamefont {{Mitra}}}, \bibinfo {author} {\bibfnamefont {M.~A.}\
  \bibnamefont {{Miville-Desch{\^e}nes}}}, \bibinfo {author} {\bibfnamefont
  {D.}~\bibnamefont {{Molinari}}}, \bibinfo {author} {\bibfnamefont
  {L.}~\bibnamefont {{Montier}}}, \bibinfo {author} {\bibfnamefont
  {G.}~\bibnamefont {{Morgante}}}, \bibinfo {author} {\bibfnamefont
  {A.}~\bibnamefont {{Moss}}}, \bibinfo {author} {\bibfnamefont
  {P.}~\bibnamefont {{Natoli}}}, \bibinfo {author} {\bibfnamefont {H.~U.}\
  \bibnamefont {{N{\o}rgaard-Nielsen}}}, \bibinfo {author} {\bibfnamefont
  {L.}~\bibnamefont {{Pagano}}}, \bibinfo {author} {\bibfnamefont
  {D.}~\bibnamefont {{Paoletti}}}, \bibinfo {author} {\bibfnamefont
  {B.}~\bibnamefont {{Partridge}}}, \bibinfo {author} {\bibfnamefont
  {G.}~\bibnamefont {{Patanchon}}}, \bibinfo {author} {\bibfnamefont {H.~V.}\
  \bibnamefont {{Peiris}}}, \bibinfo {author} {\bibfnamefont {F.}~\bibnamefont
  {{Perrotta}}}, \bibinfo {author} {\bibfnamefont {V.}~\bibnamefont
  {{Pettorino}}}, \bibinfo {author} {\bibfnamefont {F.}~\bibnamefont
  {{Piacentini}}}, \bibinfo {author} {\bibfnamefont {L.}~\bibnamefont
  {{Polastri}}}, \bibinfo {author} {\bibfnamefont {G.}~\bibnamefont
  {{Polenta}}}, \bibinfo {author} {\bibfnamefont {J.~L.}\ \bibnamefont
  {{Puget}}}, \bibinfo {author} {\bibfnamefont {J.~P.}\ \bibnamefont
  {{Rachen}}}, \bibinfo {author} {\bibfnamefont {M.}~\bibnamefont
  {{Reinecke}}}, \bibinfo {author} {\bibfnamefont {M.}~\bibnamefont
  {{Remazeilles}}}, \bibinfo {author} {\bibfnamefont {A.}~\bibnamefont
  {{Renzi}}}, \bibinfo {author} {\bibfnamefont {G.}~\bibnamefont {{Rocha}}},
  \bibinfo {author} {\bibfnamefont {C.}~\bibnamefont {{Rosset}}}, \bibinfo
  {author} {\bibfnamefont {G.}~\bibnamefont {{Roudier}}}, \bibinfo {author}
  {\bibfnamefont {J.~A.}\ \bibnamefont {{Rubi{\~n}o-Mart{\'\i}n}}}, \bibinfo
  {author} {\bibfnamefont {B.}~\bibnamefont {{Ruiz-Granados}}}, \bibinfo
  {author} {\bibfnamefont {L.}~\bibnamefont {{Salvati}}}, \bibinfo {author}
  {\bibfnamefont {M.}~\bibnamefont {{Sandri}}}, \bibinfo {author}
  {\bibfnamefont {M.}~\bibnamefont {{Savelainen}}}, \bibinfo {author}
  {\bibfnamefont {D.}~\bibnamefont {{Scott}}}, \bibinfo {author} {\bibfnamefont
  {E.~P.~S.}\ \bibnamefont {{Shellard}}}, \bibinfo {author} {\bibfnamefont
  {C.}~\bibnamefont {{Sirignano}}}, \bibinfo {author} {\bibfnamefont
  {G.}~\bibnamefont {{Sirri}}}, \bibinfo {author} {\bibfnamefont {L.~D.}\
  \bibnamefont {{Spencer}}}, \bibinfo {author} {\bibfnamefont {R.}~\bibnamefont
  {{Sunyaev}}}, \bibinfo {author} {\bibfnamefont {A.~S.}\ \bibnamefont
  {{Suur-Uski}}}, \bibinfo {author} {\bibfnamefont {J.~A.}\ \bibnamefont
  {{Tauber}}}, \bibinfo {author} {\bibfnamefont {D.}~\bibnamefont
  {{Tavagnacco}}}, \bibinfo {author} {\bibfnamefont {M.}~\bibnamefont
  {{Tenti}}}, \bibinfo {author} {\bibfnamefont {L.}~\bibnamefont
  {{Toffolatti}}}, \bibinfo {author} {\bibfnamefont {M.}~\bibnamefont
  {{Tomasi}}}, \bibinfo {author} {\bibfnamefont {T.}~\bibnamefont
  {{Trombetti}}}, \bibinfo {author} {\bibfnamefont {L.}~\bibnamefont
  {{Valenziano}}}, \bibinfo {author} {\bibfnamefont {J.}~\bibnamefont
  {{Valiviita}}}, \bibinfo {author} {\bibfnamefont {B.}~\bibnamefont {{Van
  Tent}}}, \bibinfo {author} {\bibfnamefont {L.}~\bibnamefont {{Vibert}}},
  \bibinfo {author} {\bibfnamefont {P.}~\bibnamefont {{Vielva}}}, \bibinfo
  {author} {\bibfnamefont {F.}~\bibnamefont {{Villa}}}, \bibinfo {author}
  {\bibfnamefont {N.}~\bibnamefont {{Vittorio}}}, \bibinfo {author}
  {\bibfnamefont {B.~D.}\ \bibnamefont {{Wand elt}}}, \bibinfo {author}
  {\bibfnamefont {I.~K.}\ \bibnamefont {{Wehus}}}, \bibinfo {author}
  {\bibfnamefont {M.}~\bibnamefont {{White}}}, \bibinfo {author} {\bibfnamefont
  {S.~D.~M.}\ \bibnamefont {{White}}}, \bibinfo {author} {\bibfnamefont
  {A.}~\bibnamefont {{Zacchei}}}, \ and\ \bibinfo {author} {\bibfnamefont
  {A.}~\bibnamefont {{Zonca}}},\ }\href@noop {} {\bibfield  {journal} {\bibinfo
   {journal} {arXiv e-prints}\ ,\ \bibinfo {eid} {arXiv:1807.06209}} (\bibinfo
  {year} {2018})},\ \Eprint {http://arxiv.org/abs/1807.06209} {arXiv:1807.06209
  [astro-ph.CO]} \BibitemShut {NoStop}%
\bibitem [{\citenamefont {{Sedov}}(1959)}]{sedov}%
  \BibitemOpen
  \bibfield  {author} {\bibinfo {author} {\bibfnamefont {L.~I.}\ \bibnamefont
  {{Sedov}}},\ }\href@noop {} {\emph {\bibinfo {title} {Similarity and
  Dimensional Methods in Mechanics, New York: Academic Press, 1959}}}\
  (\bibinfo {year} {1959})\BibitemShut {NoStop}%
\bibitem [{\citenamefont {Liddle}(2013)}]{liddle}%
  \BibitemOpen
  \bibfield  {author} {\bibinfo {author} {\bibfnamefont {A.}~\bibnamefont
  {Liddle}},\ }\href@noop {} {\emph {\bibinfo {title} {An Introduction to
  Modern Cosmology}}}\ (\bibinfo  {publisher} {Wiley},\ \bibinfo {year}
  {2013})\BibitemShut {NoStop}%
\bibitem [{\citenamefont {Dodelson}\ and\ \citenamefont
  {1941-1969).}(2003)}]{dodelson}%
  \BibitemOpen
  \bibfield  {author} {\bibinfo {author} {\bibfnamefont {S.}~\bibnamefont
  {Dodelson}}\ and\ \bibinfo {author} {\bibfnamefont {A.~P. L.~.}\ \bibnamefont
  {1941-1969).}},\ }\href {https://books.google.com.mx/books?id=3oPRxdXJexcC}
  {\emph {\bibinfo {title} {Modern Cosmology}}}\ (\bibinfo  {publisher}
  {Elsevier Science},\ \bibinfo {year} {2003})\BibitemShut {NoStop}%
\bibitem [{\citenamefont {{Perlmutter}}\ \emph {et~al.}(1999)\citenamefont
  {{Perlmutter}}, \citenamefont {{Aldering}}, \citenamefont {{Goldhaber}},
  \citenamefont {{Knop}}, \citenamefont {{Nugent}}, \citenamefont {{Castro}},
  \citenamefont {{Deustua}}, \citenamefont {{Fabbro}}, \citenamefont
  {{Goobar}}, \citenamefont {{Groom}}, \citenamefont {{Hook}}, \citenamefont
  {{Kim}}, \citenamefont {{Kim}}, \citenamefont {{Lee}}, \citenamefont
  {{Nunes}}, \citenamefont {{Pain}}, \citenamefont {{Pennypacker}},
  \citenamefont {{Quimby}}, \citenamefont {{Lidman}}, \citenamefont {{Ellis}},
  \citenamefont {{Irwin}}, \citenamefont {{McMahon}}, \citenamefont
  {{Ruiz-Lapuente}}, \citenamefont {{Walton}}, \citenamefont {{Schaefer}},
  \citenamefont {{Boyle}}, \citenamefont {{Filippenko}}, \citenamefont
  {{Matheson}}, \citenamefont {{Fruchter}}, \citenamefont {{Panagia}},
  \citenamefont {{Newberg}}, \citenamefont {{Couch}},\ and\ \citenamefont
  {{Project}}}]{Perlmutter99}%
  \BibitemOpen
  \bibfield  {author} {\bibinfo {author} {\bibfnamefont {S.}~\bibnamefont
  {{Perlmutter}}}, \bibinfo {author} {\bibfnamefont {G.}~\bibnamefont
  {{Aldering}}}, \bibinfo {author} {\bibfnamefont {G.}~\bibnamefont
  {{Goldhaber}}}, \bibinfo {author} {\bibfnamefont {R.~A.}\ \bibnamefont
  {{Knop}}}, \bibinfo {author} {\bibfnamefont {P.}~\bibnamefont {{Nugent}}},
  \bibinfo {author} {\bibfnamefont {P.~G.}\ \bibnamefont {{Castro}}}, \bibinfo
  {author} {\bibfnamefont {S.}~\bibnamefont {{Deustua}}}, \bibinfo {author}
  {\bibfnamefont {S.}~\bibnamefont {{Fabbro}}}, \bibinfo {author}
  {\bibfnamefont {A.}~\bibnamefont {{Goobar}}}, \bibinfo {author}
  {\bibfnamefont {D.~E.}\ \bibnamefont {{Groom}}}, \bibinfo {author}
  {\bibfnamefont {I.~M.}\ \bibnamefont {{Hook}}}, \bibinfo {author}
  {\bibfnamefont {A.~G.}\ \bibnamefont {{Kim}}}, \bibinfo {author}
  {\bibfnamefont {M.~Y.}\ \bibnamefont {{Kim}}}, \bibinfo {author}
  {\bibfnamefont {J.~C.}\ \bibnamefont {{Lee}}}, \bibinfo {author}
  {\bibfnamefont {N.~J.}\ \bibnamefont {{Nunes}}}, \bibinfo {author}
  {\bibfnamefont {R.}~\bibnamefont {{Pain}}}, \bibinfo {author} {\bibfnamefont
  {C.~R.}\ \bibnamefont {{Pennypacker}}}, \bibinfo {author} {\bibfnamefont
  {R.}~\bibnamefont {{Quimby}}}, \bibinfo {author} {\bibfnamefont
  {C.}~\bibnamefont {{Lidman}}}, \bibinfo {author} {\bibfnamefont {R.~S.}\
  \bibnamefont {{Ellis}}}, \bibinfo {author} {\bibfnamefont {M.}~\bibnamefont
  {{Irwin}}}, \bibinfo {author} {\bibfnamefont {R.~G.}\ \bibnamefont
  {{McMahon}}}, \bibinfo {author} {\bibfnamefont {P.}~\bibnamefont
  {{Ruiz-Lapuente}}}, \bibinfo {author} {\bibfnamefont {N.}~\bibnamefont
  {{Walton}}}, \bibinfo {author} {\bibfnamefont {B.}~\bibnamefont
  {{Schaefer}}}, \bibinfo {author} {\bibfnamefont {B.~J.}\ \bibnamefont
  {{Boyle}}}, \bibinfo {author} {\bibfnamefont {A.~V.}\ \bibnamefont
  {{Filippenko}}}, \bibinfo {author} {\bibfnamefont {T.}~\bibnamefont
  {{Matheson}}}, \bibinfo {author} {\bibfnamefont {A.~S.}\ \bibnamefont
  {{Fruchter}}}, \bibinfo {author} {\bibfnamefont {N.}~\bibnamefont
  {{Panagia}}}, \bibinfo {author} {\bibfnamefont {H.~J.~M.}\ \bibnamefont
  {{Newberg}}}, \bibinfo {author} {\bibfnamefont {W.~J.}\ \bibnamefont
  {{Couch}}}, \ and\ \bibinfo {author} {\bibfnamefont {T.~S.~C.}\ \bibnamefont
  {{Project}}},\ }\href {\doibase 10.1086/307221} {\bibfield  {journal}
  {\bibinfo  {journal} {\apj}\ }\textbf {\bibinfo {volume} {517}},\ \bibinfo
  {pages} {565} (\bibinfo {year} {1999})},\ \Eprint
  {http://arxiv.org/abs/astro-ph/9812133} {arXiv:astro-ph/9812133 [astro-ph]}
  \BibitemShut {NoStop}%
\bibitem [{\citenamefont {Visser}(2005)}]{Visser}%
  \BibitemOpen
  \bibfield  {author} {\bibinfo {author} {\bibfnamefont {M.}~\bibnamefont
  {Visser}},\ }\href {\doibase 10.1007/s10714-005-0134-8} {\bibfield  {journal}
  {\bibinfo  {journal} {Gen. Rel. Grav.}\ }\textbf {\bibinfo {volume} {37}},\
  \bibinfo {pages} {1541} (\bibinfo {year} {2005})},\ \Eprint
  {http://arxiv.org/abs/gr-qc/0411131} {arXiv:gr-qc/0411131} \BibitemShut
  {NoStop}%
\bibitem [{\citenamefont {{Suzuki}}\ \emph {et~al.}(2012)\citenamefont
  {{Suzuki}}, \citenamefont {{Rubin}}, \citenamefont {{Lidman}}, \citenamefont
  {{Aldering}}, \citenamefont {{Amanullah}}, \citenamefont {{Barbary}},
  \citenamefont {{Barrientos}}, \citenamefont {{Botyanszki}}, \citenamefont
  {{Brodwin}}, \citenamefont {{Connolly}}, \citenamefont {{Dawson}},
  \citenamefont {{Dey}}, \citenamefont {{Doi}}, \citenamefont {{Donahue}},
  \citenamefont {{Deustua}}, \citenamefont {{Eisenhardt}}, \citenamefont
  {{Ellingson}}, \citenamefont {{Faccioli}}, \citenamefont {{Fadeyev}},
  \citenamefont {{Fakhouri}}, \citenamefont {{Fruchter}}, \citenamefont
  {{Gilbank}}, \citenamefont {{Gladders}}, \citenamefont {{Goldhaber}},
  \citenamefont {{Gonzalez}}, \citenamefont {{Goobar}}, \citenamefont {{Gude}},
  \citenamefont {{Hattori}}, \citenamefont {{Hoekstra}}, \citenamefont
  {{Hsiao}}, \citenamefont {{Huang}}, \citenamefont {{Ihara}}, \citenamefont
  {{Jee}}, \citenamefont {{Johnston}}, \citenamefont {{Kashikawa}},
  \citenamefont {{Koester}}, \citenamefont {{Konishi}}, \citenamefont
  {{Kowalski}}, \citenamefont {{Linder}}, \citenamefont {{Lubin}},
  \citenamefont {{Melbourne}}, \citenamefont {{Meyers}}, \citenamefont
  {{Morokuma}}, \citenamefont {{Munshi}}, \citenamefont {{Mullis}},
  \citenamefont {{Oda}}, \citenamefont {{Panagia}}, \citenamefont
  {{Perlmutter}}, \citenamefont {{Postman}}, \citenamefont {{Pritchard}},
  \citenamefont {{Rhodes}}, \citenamefont {{Ripoche}}, \citenamefont
  {{Rosati}}, \citenamefont {{Schlegel}}, \citenamefont {{Spadafora}},
  \citenamefont {{Stanford}}, \citenamefont {{Stanishev}}, \citenamefont
  {{Stern}}, \citenamefont {{Strovink}}, \citenamefont {{Takanashi}},
  \citenamefont {{Tokita}}, \citenamefont {{Wagner}}, \citenamefont {{Wang}},
  \citenamefont {{Yasuda}}, \citenamefont {{Yee}},\ and\ \citenamefont
  {{Supernova Cosmology Project}}}]{SCP-Union2012}%
  \BibitemOpen
  \bibfield  {author} {\bibinfo {author} {\bibfnamefont {N.}~\bibnamefont
  {{Suzuki}}}, \bibinfo {author} {\bibfnamefont {D.}~\bibnamefont {{Rubin}}},
  \bibinfo {author} {\bibfnamefont {C.}~\bibnamefont {{Lidman}}}, \bibinfo
  {author} {\bibfnamefont {G.}~\bibnamefont {{Aldering}}}, \bibinfo {author}
  {\bibfnamefont {R.}~\bibnamefont {{Amanullah}}}, \bibinfo {author}
  {\bibfnamefont {K.}~\bibnamefont {{Barbary}}}, \bibinfo {author}
  {\bibfnamefont {L.~F.}\ \bibnamefont {{Barrientos}}}, \bibinfo {author}
  {\bibfnamefont {J.}~\bibnamefont {{Botyanszki}}}, \bibinfo {author}
  {\bibfnamefont {M.}~\bibnamefont {{Brodwin}}}, \bibinfo {author}
  {\bibfnamefont {N.}~\bibnamefont {{Connolly}}}, \bibinfo {author}
  {\bibfnamefont {K.~S.}\ \bibnamefont {{Dawson}}}, \bibinfo {author}
  {\bibfnamefont {A.}~\bibnamefont {{Dey}}}, \bibinfo {author} {\bibfnamefont
  {M.}~\bibnamefont {{Doi}}}, \bibinfo {author} {\bibfnamefont
  {M.}~\bibnamefont {{Donahue}}}, \bibinfo {author} {\bibfnamefont
  {S.}~\bibnamefont {{Deustua}}}, \bibinfo {author} {\bibfnamefont
  {P.}~\bibnamefont {{Eisenhardt}}}, \bibinfo {author} {\bibfnamefont
  {E.}~\bibnamefont {{Ellingson}}}, \bibinfo {author} {\bibfnamefont
  {L.}~\bibnamefont {{Faccioli}}}, \bibinfo {author} {\bibfnamefont
  {V.}~\bibnamefont {{Fadeyev}}}, \bibinfo {author} {\bibfnamefont {H.~K.}\
  \bibnamefont {{Fakhouri}}}, \bibinfo {author} {\bibfnamefont {A.~S.}\
  \bibnamefont {{Fruchter}}}, \bibinfo {author} {\bibfnamefont {D.~G.}\
  \bibnamefont {{Gilbank}}}, \bibinfo {author} {\bibfnamefont {M.~D.}\
  \bibnamefont {{Gladders}}}, \bibinfo {author} {\bibfnamefont
  {G.}~\bibnamefont {{Goldhaber}}}, \bibinfo {author} {\bibfnamefont {A.~H.}\
  \bibnamefont {{Gonzalez}}}, \bibinfo {author} {\bibfnamefont
  {A.}~\bibnamefont {{Goobar}}}, \bibinfo {author} {\bibfnamefont
  {A.}~\bibnamefont {{Gude}}}, \bibinfo {author} {\bibfnamefont
  {T.}~\bibnamefont {{Hattori}}}, \bibinfo {author} {\bibfnamefont
  {H.}~\bibnamefont {{Hoekstra}}}, \bibinfo {author} {\bibfnamefont
  {E.}~\bibnamefont {{Hsiao}}}, \bibinfo {author} {\bibfnamefont
  {X.}~\bibnamefont {{Huang}}}, \bibinfo {author} {\bibfnamefont
  {Y.}~\bibnamefont {{Ihara}}}, \bibinfo {author} {\bibfnamefont {M.~J.}\
  \bibnamefont {{Jee}}}, \bibinfo {author} {\bibfnamefont {D.}~\bibnamefont
  {{Johnston}}}, \bibinfo {author} {\bibfnamefont {N.}~\bibnamefont
  {{Kashikawa}}}, \bibinfo {author} {\bibfnamefont {B.}~\bibnamefont
  {{Koester}}}, \bibinfo {author} {\bibfnamefont {K.}~\bibnamefont
  {{Konishi}}}, \bibinfo {author} {\bibfnamefont {M.}~\bibnamefont
  {{Kowalski}}}, \bibinfo {author} {\bibfnamefont {E.~V.}\ \bibnamefont
  {{Linder}}}, \bibinfo {author} {\bibfnamefont {L.}~\bibnamefont {{Lubin}}},
  \bibinfo {author} {\bibfnamefont {J.}~\bibnamefont {{Melbourne}}}, \bibinfo
  {author} {\bibfnamefont {J.}~\bibnamefont {{Meyers}}}, \bibinfo {author}
  {\bibfnamefont {T.}~\bibnamefont {{Morokuma}}}, \bibinfo {author}
  {\bibfnamefont {F.}~\bibnamefont {{Munshi}}}, \bibinfo {author}
  {\bibfnamefont {C.}~\bibnamefont {{Mullis}}}, \bibinfo {author}
  {\bibfnamefont {T.}~\bibnamefont {{Oda}}}, \bibinfo {author} {\bibfnamefont
  {N.}~\bibnamefont {{Panagia}}}, \bibinfo {author} {\bibfnamefont
  {S.}~\bibnamefont {{Perlmutter}}}, \bibinfo {author} {\bibfnamefont
  {M.}~\bibnamefont {{Postman}}}, \bibinfo {author} {\bibfnamefont
  {T.}~\bibnamefont {{Pritchard}}}, \bibinfo {author} {\bibfnamefont
  {J.}~\bibnamefont {{Rhodes}}}, \bibinfo {author} {\bibfnamefont
  {P.}~\bibnamefont {{Ripoche}}}, \bibinfo {author} {\bibfnamefont
  {P.}~\bibnamefont {{Rosati}}}, \bibinfo {author} {\bibfnamefont {D.~J.}\
  \bibnamefont {{Schlegel}}}, \bibinfo {author} {\bibfnamefont
  {A.}~\bibnamefont {{Spadafora}}}, \bibinfo {author} {\bibfnamefont {S.~A.}\
  \bibnamefont {{Stanford}}}, \bibinfo {author} {\bibfnamefont
  {V.}~\bibnamefont {{Stanishev}}}, \bibinfo {author} {\bibfnamefont
  {D.}~\bibnamefont {{Stern}}}, \bibinfo {author} {\bibfnamefont
  {M.}~\bibnamefont {{Strovink}}}, \bibinfo {author} {\bibfnamefont
  {N.}~\bibnamefont {{Takanashi}}}, \bibinfo {author} {\bibfnamefont
  {K.}~\bibnamefont {{Tokita}}}, \bibinfo {author} {\bibfnamefont
  {M.}~\bibnamefont {{Wagner}}}, \bibinfo {author} {\bibfnamefont
  {L.}~\bibnamefont {{Wang}}}, \bibinfo {author} {\bibfnamefont
  {N.}~\bibnamefont {{Yasuda}}}, \bibinfo {author} {\bibfnamefont {H.~K.~C.}\
  \bibnamefont {{Yee}}}, \ and\ \bibinfo {author} {\bibfnamefont
  {T.}~\bibnamefont {{Supernova Cosmology Project}}},\ }\href {\doibase
  10.1088/0004-637X/746/1/85} {\bibfield  {journal} {\bibinfo  {journal}
  {\apj}\ }\textbf {\bibinfo {volume} {746}},\ \bibinfo {eid} {85} (\bibinfo
  {year} {2012})},\ \Eprint {http://arxiv.org/abs/1105.3470} {arXiv:1105.3470
  [astro-ph.CO]} \BibitemShut {NoStop}%
\bibitem [{\citenamefont {{Giusti}}(2020)}]{andrea}%
  \BibitemOpen
  \bibfield  {author} {\bibinfo {author} {\bibfnamefont {A.}~\bibnamefont
  {{Giusti}}},\ }\href {\doibase 10.1103/PhysRevD.101.124029} {\bibfield
  {journal} {\bibinfo  {journal} {\prd}\ }\textbf {\bibinfo {volume} {101}},\
  \bibinfo {eid} {124029} (\bibinfo {year} {2020})},\ \Eprint
  {http://arxiv.org/abs/2002.07133} {arXiv:2002.07133 [gr-qc]} \BibitemShut
  {NoStop}%
\bibitem [{\citenamefont {{Milgrom}}(1983)}]{Milgrom1}%
  \BibitemOpen
  \bibfield  {author} {\bibinfo {author} {\bibfnamefont {M.}~\bibnamefont
  {{Milgrom}}},\ }\href {\doibase 10.1086/161130} {\bibfield  {journal}
  {\bibinfo  {journal} {\apj}\ }\textbf {\bibinfo {volume} {270}},\ \bibinfo
  {pages} {365} (\bibinfo {year} {1983})}\BibitemShut {NoStop}%
\bibitem [{\citenamefont {{Milgrom}}(2008)}]{Milgrom:2008}%
  \BibitemOpen
  \bibfield  {author} {\bibinfo {author} {\bibfnamefont {M.}~\bibnamefont
  {{Milgrom}}},\ }\href@noop {} {\bibfield  {journal} {\bibinfo  {journal}
  {arXiv e-prints}\ ,\ \bibinfo {eid} {arXiv:0801.3133}} (\bibinfo {year}
  {2008})},\ \Eprint {http://arxiv.org/abs/0801.3133} {arXiv:0801.3133
  [astro-ph]} \BibitemShut {NoStop}%
\bibitem [{\citenamefont {{Bernal}}\ \emph
  {et~al.}(2011{\natexlab{b}})\citenamefont {{Bernal}}, \citenamefont
  {{Capozziello}}, \citenamefont {{Cristofano}},\ and\ \citenamefont {{de
  Laurentis}}}]{Bernal:2011a}%
  \BibitemOpen
  \bibfield  {author} {\bibinfo {author} {\bibfnamefont {T.}~\bibnamefont
  {{Bernal}}}, \bibinfo {author} {\bibfnamefont {S.}~\bibnamefont
  {{Capozziello}}}, \bibinfo {author} {\bibfnamefont {G.}~\bibnamefont
  {{Cristofano}}}, \ and\ \bibinfo {author} {\bibfnamefont {M.}~\bibnamefont
  {{de Laurentis}}},\ }\href {\doibase 10.1142/S0217732311037042} {\bibfield
  {journal} {\bibinfo  {journal} {Modern Physics Letters A}\ }\textbf {\bibinfo
  {volume} {26}},\ \bibinfo {pages} {2677} (\bibinfo {year}
  {2011}{\natexlab{b}})},\ \Eprint {http://arxiv.org/abs/1110.2580}
  {arXiv:1110.2580 [gr-qc]} \BibitemShut {NoStop}%
\bibitem [{\citenamefont {Oldham}\ and\ \citenamefont
  {Spanier}(2006)}]{oldham}%
  \BibitemOpen
  \bibfield  {author} {\bibinfo {author} {\bibfnamefont {K.}~\bibnamefont
  {Oldham}}\ and\ \bibinfo {author} {\bibfnamefont {J.}~\bibnamefont
  {Spanier}},\ }\href {https://books.google.com.mx/books?id=yh68AAAACAAJ}
  {\emph {\bibinfo {title} {The Fractional Calculus: Theory and Applications of
  Differentiation and Integration to Arbitrary Order}}},\ Dover books on
  mathematics\ (\bibinfo  {publisher} {Dover Publications},\ \bibinfo {year}
  {2006})\BibitemShut {NoStop}%
\bibitem [{\citenamefont {Camarena}\ and\ \citenamefont
  {Marra}(2020)}]{camarena}%
  \BibitemOpen
  \bibfield  {author} {\bibinfo {author} {\bibfnamefont {D.}~\bibnamefont
  {Camarena}}\ and\ \bibinfo {author} {\bibfnamefont {V.}~\bibnamefont
  {Marra}},\ }\href {\doibase 10.1103/PhysRevResearch.2.013028} {\bibfield
  {journal} {\bibinfo  {journal} {Phys. Rev. Research}\ }\textbf {\bibinfo
  {volume} {2}},\ \bibinfo {pages} {013028} (\bibinfo {year}
  {2020})}\BibitemShut {NoStop}%
\bibitem [{\citenamefont {{Mukherjee}}\ \emph {et~al.}(2020)\citenamefont
  {{Mukherjee}}, \citenamefont {{Ghosh}}, \citenamefont {{Graham}},
  \citenamefont {{Karathanasis}}, \citenamefont {{Kasliwal}}, \citenamefont
  {{Maga{\~n}a Hernandez}}, \citenamefont {{Nissanke}}, \citenamefont
  {{Silvestri}},\ and\ \citenamefont {{Wandelt}}}]{Mukherjee}%
  \BibitemOpen
  \bibfield  {author} {\bibinfo {author} {\bibfnamefont {S.}~\bibnamefont
  {{Mukherjee}}}, \bibinfo {author} {\bibfnamefont {A.}~\bibnamefont
  {{Ghosh}}}, \bibinfo {author} {\bibfnamefont {M.~J.}\ \bibnamefont
  {{Graham}}}, \bibinfo {author} {\bibfnamefont {C.}~\bibnamefont
  {{Karathanasis}}}, \bibinfo {author} {\bibfnamefont {M.~M.}\ \bibnamefont
  {{Kasliwal}}}, \bibinfo {author} {\bibfnamefont {I.}~\bibnamefont
  {{Maga{\~n}a Hernandez}}}, \bibinfo {author} {\bibfnamefont {S.~M.}\
  \bibnamefont {{Nissanke}}}, \bibinfo {author} {\bibfnamefont
  {A.}~\bibnamefont {{Silvestri}}}, \ and\ \bibinfo {author} {\bibfnamefont
  {B.~D.}\ \bibnamefont {{Wandelt}}},\ }\href@noop {} {\bibfield  {journal}
  {\bibinfo  {journal} {arXiv e-prints}\ ,\ \bibinfo {eid} {arXiv:2009.14199}}
  (\bibinfo {year} {2020})},\ \Eprint {http://arxiv.org/abs/2009.14199}
  {arXiv:2009.14199 [astro-ph.CO]} \BibitemShut {NoStop}%
\end{thebibliography}%
